\immediate\write18{bibtex calco2019-lmcs}
\documentclass{lmcs}
\usepackage[utf8]{inputenc}
\pdfoutput=1

\usepackage{lastpage}
\lmcsdoi{19}{1}{8}
\lmcsheading{}{\pageref{LastPage}}{}{}%
{Apr.~21,~2020}{Jan.~26,~2023}{}

\usepackage{hyperref}

\usepackage{mathtools}

\theoremstyle{definition}
\newtheorem{notation}[thm]{Notation}

\newcommand{\sembox}[3]{\langle\!\langle#1]\!]#2[\![#3\rangle\!\rangle} % chktex 9
\newcommand{\semdia}[3]{[\![#1\rangle\!\rangle#2\langle\!\langle#3]\!]}

\newcommand\concat{\mathbin{+\mkern-10mu+}}
\newcommand\eqdef{\stackrel{\mathclap{\tiny\mbox{def}}}{=}}

\newcommand{\hcomp}{\hspace{0.24ex};}
\newcommand{\names}{\mathcal N}

\newcommand{\id}{\mathit{id}}

\newcommand{\Nom}{\mathsf{Nom}}
\newcommand{\supp}{\mathsf{supp}}

\newcommand{\Fun}{\deflink{ex:nF}{\mathsf n\mathbb F} }

\newcommand{\nF}{\deflink{ex:nF}{\mathsf n\mathbb F} }
\newcommand{\FunSkel}{\deflink{funSkel-def}{\mathbb F} }

\newcommand{\Cat}{\mathsf{Cat}}
\newcommand{\V}{\mathcal{V}}

\newcommand{\PRO}{\mathsf{PRO}}
\newcommand{\then}{{;}}
\newcommand{\dom}{\ensuremath{\mathbf{dom}}}
\newcommand{\cod}{\ensuremath{\mathbf{cod}}}
\newcommand{\comp}{\ensuremath{\mathbf{comp}}}
\newcommand{\compl}{\ensuremath{\mathbf{compl}}}
\newcommand{\compr}{\ensuremath{\mathbf{compr}}}
\newcommand{\leftt}{\ensuremath{\mathbf{left}}}
\newcommand{\rightt}{\ensuremath{\mathbf{right}}}

\newcommand{\PROP}{\deflink{PROP-def}{\mathsf{PROP}}}
\newcommand{\nPROP}{\deflink{def:nprop}{\mathsf{nPROP}}}
\newcommand{\Prop}{ \deflink{def:Prop-def}{\mathsf{Prop}} }
\newcommand{\nProp}{ \deflink{def:nProp-def}{\mathsf{nProp}} }
\newcommand{\nfSmt}{\deflink{nfSmt-def}{\mathsf{nfSmt}}}
\newcommand{\nfSmtSubscript}{\deflink{nfSmt-def}{\mathsf{nfSmt_{}}}}
\newcommand{\nfNmt}{\deflink{nfNmt-def}{\mathsf{nfNmt}}}
\newcommand{\nfNmtSubscript}{\deflink{nfNmt-def}{\mathsf{nfNmt_{}}}}
\newcommand{\injNmt}{\deflink{injNmt-def}{\,\iota\,}}
\newcommand{\injNmtSubscript}{\deflink{injNmt-def}{\,\iota\,_{}}}
\newcommand{\Nmt}{\deflink{Nmt-def}{\mathsf{Nmt}}}
\newcommand{\NmtSubscript}{\deflink{Nmt-def}{\mathsf{Nmt_{}}}}
\newcommand{\NMT}{\deflink{NMT-def}{\mathsf{NMT}}}
\newcommand{\NMTeq}{\deflink{NMTeq}{\mathbf{NMT}}}
\newcommand{\eqNMT}{\deflink{NMTeq}{\,\stackrel{\mathclap{\tiny\mbox{NMT}}}{=}\,}}
\newcommand{\eqSMT}{\deflink{SMTeq}{\,\stackrel{\mathclap{\tiny\mbox{SMT}}}{=}\,}}
\newcommand{\injSmt}{\deflink{injSmt-def}{\,\iota\,}}
\newcommand{\injSmtSubscript}{\deflink{injSmt-def}{\,\iota\,_{}}}
\newcommand{\Smt}{\deflink{Smt-def}{\mathsf{Smt}}}
\newcommand{\SmtSubscript}{\deflink{Smt-def}{\mathsf{Smt_{}}}}
\newcommand{\SMT}{\deflink{SMT-def}{\mathsf{SMT}}}
\newcommand{\SMTeq}{\deflink{SMTeq}{\mathbf{SMT}}}
\newcommand{\Trm}{\deflink{fig:smt-terms}{\mathsf{Trm}}}
\newcommand{\boxTrm}{\deflink{box-def}{\mathrm{box}}}
\newcommand{\diaTrm}{\deflink{dia-def}{\mathrm{dia}}}
\newcommand{\nTrm}{\deflink{fig:terms}{\mathsf{nTrm}}}
\DeclareFontFamily{OT1}{pzc}{}
\DeclareFontShape{OT1}{pzc}{m}{it}{<-> s * [1.10] pzcmi7t}{}
\DeclareMathAlphabet{\mathcalligra}{OT1}{pzc}{m}{it}

\newcommand{\Cl}{\deflink{fig:cl-op}{\mathcalligra{Th}}}
\newcommand{\ORD}{\deflink{prop:ORD}{\mathsf{ORD}}}
\newcommand{\ORDSubscript}{\deflink{prop:ORD}{\mathsf{ORD_{}}}}
\newcommand{\ORDeq}{\deflink{prop:ORD}{\mathbf{ORD}}}
\newcommand{\eqORD}{\,\deflink{eqORD-def}{\, \stackrel{\mathclap{\tiny\mbox{ORD}}}{=} \,}\,}
\newcommand{\NOM}{\deflink{prop:NOM}{\mathsf{NOM}}}
\newcommand{\NOMSubscript}{\deflink{prop:NOM}{\mathsf{NOM_{}}}}
\newcommand{\NOMeq}{\deflink{prop:NOM}{\mathbf{NOM}}}

\newcommand{\eqNOM}{\,\deflink{eqNOM-def}{\, \stackrel{\mathclap{\tiny\mbox{NOM}}}{=} \,}\,}

\newcommand{\tw}{\sigma}
\newcommand{\gen}{\gamma}

\newcommand{\bijbox}{\colorbox{black!30}{\strut bijections $\mathsf B$}}
\newcommand{\injbox}{\colorbox{orange!60}{\strut injections $\mathsf I$}}
\newcommand{\surjbox}{\colorbox{green!40}{\strut surjections $\mathsf S$}}
\newcommand{\funbox}{\colorbox{cyan!40}{\strut functions $\mathsf F$}}
\newcommand{\pfunbox}{\colorbox{magenta!40}{\strut partial functions $\mathsf P$}}
\newcommand{\relbox}{\colorbox{cyan!60!magenta!60}{\strut relations $\mathsf R$}}
\newcommand{\mmapbox}{\colorbox{yellow!60}{\strut monotone maps $\mathsf{M}$}}
\newcommand{\bijnbox}{\colorbox{black!30}{\strut bijections $\mathsf n\mathsf B$}}
\newcommand{\injnbox}{\colorbox{orange!60}{\strut injections $\mathsf n\mathsf I$}}
\newcommand{\surjnbox}{\colorbox{green!40}{\strut surjections $\mathsf n\mathsf S$}}
\newcommand{\funnbox}{\colorbox{cyan!40}{\strut functions $\mathsf n\mathsf F$}}
\newcommand{\pfunnbox}{\colorbox{magenta!40}{\strut partial functions $\mathsf n\mathsf P$}}
\newcommand{\relnbox}{\colorbox{cyan!60!magenta!60}{\strut relations $\mathsf n\mathsf R$}}

\definecolor{darkblue}{RGB}{34, 76, 152}
\newcommand{\deflink}[2]{\hyperref[#1]{\color{brown} #2}}

\usepackage{stackengine}
\newcommand\sbullet[1][.5]{\mathbin{\vcenter{\hbox{\scalebox{#1}{$\bullet$}}}}}

\newcommand{\last}{\mathrel{\ensurestackMath{\stackengine{0.5pt}{\sbullet[1.75]}{\color{white}\boldsymbol{\ast}}{O}{c}{F}{F}{L}}}}

\newcommand{\lotimes}{\mathrel{\ensurestackMath{\stackengine{0.5pt}{\sbullet[1.75]}{\color{white}\boldsymbol{\times}}{O}{c}{F}{F}{L}}}}

\usepackage{amssymb}

\let\emptyset\varnothing

\newcommand{\nSigma}{{\mathsf n\Sigma}}

\usepackage{bbold} % used for \mathbb 1
\usepackage{xcolor} % used for color!60
\usepackage[all]{xy}
\usepackage{tikz}
\usetikzlibrary{matrix,arrows,positioning}
\usepackage[export]{adjustbox} % used for \includegraphics[center]
\usepackage{environ}
\usepackage[many]{tcolorbox}
\usetikzlibrary{patterns}

\SelectTips{lu}{10 scaled 1500}
\newdir{|>}{{}*!/-2pt/:(1,+1.9)@^{|}*!/-2pt/:(1,-1.9)@_{|}}
\newdir{ |}{{\rule{0.7pt}{1.2mm}}}
\newdir{ /}{{\rule{1.2mm}{0.7pt}}}

\usepackage{stmaryrd}
\usepackage{multicol}

\newif\iflongversion
\longversiontrue

\NewEnviron{longVersion}{\iflongversion\expandafter\BODY\fi}

\newif\ifcomments
\commentstrue

\NewEnviron{ak}{\ifcomments\expandafter\begin{color}{magenta}\BODY\end{color}\fi}
\NewEnviron{sam}{\ifcomments\expandafter\begin{color}{cyan}\BODY\end{color}\fi}
\NewEnviron{gray}{\iflongversion\expandafter\begin{color}{gray}\BODY\end{color}\fi}

\newtcbox{\stripbox}{
  enhanced,nobeforeafter,
  frame code={},
  tcbox raise base,
  boxrule=0.4pt,top=0mm,bottom=0mm,right=0mm,left=0mm,
  interior code={
    \path[
      draw=gray!40,
      pattern=north east lines,
      pattern color=gray!40
    ]
    (interior.south east) rectangle (interior.north west);
  }
}

\usepackage{float}
\floatstyle{boxed}
\restylefloat{figure}

\usepackage{etoolbox}

\usepackage[normalem]{ulem}

\allowdisplaybreaks

\begin{document}
%%%%%%%%%%%%%%%%%%%%%%%%%%%%%%%%%%
%%%%%%%%%%%%%%%%%%%%%%%%%%%%%%%%%%

\title{Completeness of Nominal PROPs}

\author[S.~Balco]{Samuel Balco\lmcsorcid{0000-0001-9838-3908}\rsuper{a}}
\address{Runtime Verification Inc.}
\email{sam@definitelynotspam.email}

\author[A.~Kurz]{Alexander Kurz\lmcsorcid{0000-0002-8685-5207}\rsuper{b}}
\address{Chapman University, Orange California, USA}
\email{akurz@chapman.edu}

\maketitle

\begin{abstract}
We introduce nominal string diagrams as string diagrams internal in the category of nominal sets. This leads us to define nominal PROPs and nominal monoidal theories. We show that the categories of ordinary PROPs and nominal PROPs are equivalent. This equivalence is then extended to symmetric monoidal theories and nominal monoidal theories, which allows us to transfer completeness results between ordinary and nominal calculi for string diagrams.
\end{abstract}

%\begin{gray}
{\footnotesize
\tableofcontents
}
%\end{gray}

\newpage

%%%%%%%%%%%%
% !TEX root =  calco2019-lmcs.tex

\section{Introduction}

The purpose of this work is to give rigorous foundations to string diagrams with named wires.
To achieve this, we follow the wide-spread practice of modeling string diagrams as \textbf{pro}duct and \textbf{p}ermutation categories, or PROPs
for short. The novel feature of our approach is to internalise PROPs in the category $\Nom$ of nominal sets.
We then show that the category of ordinary  PROPs is equivalent to the category of PROPs internal in $\Nom$, or, briefly, that ordinary  PROPs are equivalent to nominal  PROPs.

This formalises an observation familiar to users of string diagrams, namely that we can afford ourselves the practical convenience of named wires without departing from the mathematical convenience of canonically ordered wires.
To illustrate this consider the following picture.
\begin{center}
\includegraphics[page=45, width=\textwidth]{twists_new}
\end{center}

On the left, we see equations between ordinary string diagrams. Sequential composition of two diagrams connects the wires respecting their order.

For example, the top left most diagram is actually a composition of several smaller diagrams:

\begin{center}
\includegraphics[page=48, width=3.5cm]{twists_new_thesis} \
\end{center}

Mathematically, a collection of wires is an ordinal. On the right we see nominal string diagrams. A collection of wires  is a set of `names'. Sequential composition connects wires that have the same name.

\begin{center}
\includegraphics[page=71, width=7cm]{twists_new_thesis} \
\end{center}

Let us look at the uppermost equations (on grey background). For ordinary string diagrams on the left, we need axioms that equate all diagrams that represent the same permutations, since the order of wires is important. On the right, we see that we do not have any equations with wire crossings. This reflects that invariance under permutations for nominal string diagrams is inherited from nominal sets. Intuitively, while in the ordinary case wires are lists, in the nominal case wires are sets.

The third row (on green background) contains the equations stating that the binary operation in question (we may call it a ``cup'') is commutative and associative and interacts in the expected way with wire-crossings (on the left) and with renamings (on the right).

Apart from proving that the PROPs defined by the equations on the left are equivalent to the nominal PROPs axiomatised by the equations on the right, we are also interested in a procedure translating ordinary equations into nominal ones, and vice versa. This needs some care as the equations involving wire-crossings (on the left) and renamings (on the right) look quite different.

Our main results are transfer theorems that allow us not only to translate a complete symmetric monoidal theory (as illustrated on the left) to a nominal monoidal theory (as shown on the right), but also to transfer completeness. We also show the other direction from the nominal to the ordinary side.

As a result of our study we obtain the equivalence of symmetric and nominal monoidal theories and can now use named wires in a completely rigorous way, transerring results back and forth. We will remark on possible applications of  nominal PROPs in the conclusions.

This is an extended version of~\cite{bk19}. In particular, Sections~\ref{sec:nfNmt}-\ref{sec:SMTcompleteness} are new.

\textbf{Hyper-links. } Due to its syntactic nature, this paper introduces a large number of technical definitions and notations, for example: $\PROP$, $\nPROP$, $\SMT$, $\NMT$, $\FunSkel$, $\Fun$, $\Trm$, $\nTrm$, $\Prop$, $\Cl$, $\NOM$, $\ORD$, $\boxTrm$, $\diaTrm$, $\NOMeq$, $\ORDeq$, $\nfNmt$,
$\nfSmt$. To make the paper easier to read, these notions are hyper-linked in the electronic version.

\textbf{Acknowledgments. }
We started to work on this project inspired by Pawel Sobocinski's course on Graphical Linear Algebra at MGS in Leicester 2016. Over the years we profited from discussions with Fredrik Dahlqvist, Giuseppe Greco, Bart Jacobs, Peter Jipsen, Samuel Mimram, Drew Moshier, Alessandra Palmigiano, David Pym, Mike Shulman, Georg Struth, Apostolos Tzimoulis and Fabio Zanasi who all influenced the paper in some way. Special thanks go to Chad Nester for pointing out the important paper by Blute et al~\cite{Blute96}, which we had missed, and to Thomas Streicher who helped  us with questions on fibrations and internal categories.

%%%%%%%%%%%%%%%%%%%%%%%%%%%%%%%%%%%%
\section{Related Work}
%%%%%%%%%%%%%%%%%%%%%%%%%%%%%%%%%%%%

We divide discussion of related work into string diagrams and nominal sets.

\textbf{String Diagrams. }
Whilst somewhat difficult to tell with certainty, arguably the first
formal definition of string diagrams appears in the habilitation thesis
of G\"unter Hotz~\cite{hotz65}. However, forms of diagrammatic reasoning
in areas such as knot theory have much earlier origins (see~\cite{przytycki98} for a nice historical summary). Definitions of string
diagrams have also been introduced, amongst others, by Penrose~\cite{penrose71}, Joyal \& Street~\cite{joyal1993,
joyal-street:tensor1} and have cropped up in presentations of sequent
calculi~\cite{Cockett18}, linear logic as proof nets~\cite{Girard87,
Mellies06}, bigraphs~\cite{Milner06}, signal flow diagrams in control
theory~\cite{Mason53} and network theory~\cite{Bonchi15} as well as in
areas such as quantum physics and computing~\cite{coecke-kissinger}.

All of these formalisms are underpinned by the same category theory,
namely that of (symmetric) monoidal categories, specifically
\textbf{pro}duct and \textbf{p}ermutation categories called \(\PROP\)s
for short, introduced by MacLane~\cite{maclane:prop}. For an overview of
classic/single sorted string diagrams see~\cite{Selinger10}.

Whilst our work is novel in its presentation of nominal string diagrams
as monoidal categories internal in \(\Nom\), we are by no means the
first to generalise \(\PROP\)s to a multi-sorted or nominal settings.
Indeed, even (one of) the earliest papers on string diagrams, namely
that of Roger Penrose~\cite{penrose71}, already introduces ``nominal''
string diagrams where the wires of his pictures are given labels.
Amongst later works, a commonly seen variation to ordinary string
diagrams is the notion of colored props~\cite{Hackney14,
ZanasiFreeHypegraph}. This generalisation from one-sorted to many-sorted $\PROP$s is orthogonal to our generalisation to nominal $\PROP$s.

Finally, we must mention the work of Blute et al.~\cite{Blute96}, which
is similar in many aspects to our work, especially in the use of the
diamond notation \(\langle-]-[-\rangle\), which we arrived at independently from the
authors. We also add the converse $[-\rangle-\langle -]$  and work out the axioms under which they are inverse to each other.

Another paper in similar
spirit, by Ghica and Lopez~\cite{ghica-lopez}, introduces a version of
nominal string diagrams by explicitly introducing names and binders for
ordinary string diagrams.

\textbf{Nominal Sets. }
Nominal sets were introduced by Gabbay and Pitts in~\cite{gabb-pitt:lics99} and are are based on Fraenkel-Mostowski set theory in which sets are equipped with an additional permutation action. Equivalent formulations based on presheaves~\cite{Hof99,FiorePT99} and named sets~\cite{MP98} were introduced around same time.

All three approaches have their own advantages. The presheaf approach explicitly types terms by their sets of free names and recognises quantifiers as adjoints as in Lawvere's hyperdoctrines~\cite{lawvere-adjoint-foundations}. Named sets give finite presentations to infinite orbits and have proved to be a suitable basis for implementing generalisations of algorithms known from automata such as partition refinement~\cite{DMT05} or Angluin's algorithm~\cite{MSSKS17}.

Nominal sets are designed to make minimal modifications to ordinary set theory and are thus well suited to internalising established areas of mathematics. Gabbay and Pitts showed that $\lambda$-calculus terms up to $\alpha$-equivalence form a term-algebra not in sets but in nominal sets. Since names and name binding play a fundamental role in programming languages, several areas of theoretical computer science have been internalised in nominal sets as witnessed for example by work on % term-rewriting~\cite{FG07}, logic programming~\cite{CheneyU08}, theorem proving~\cite{Urban08},
universal algebra~\cite{GM09,petrisan:phd}, domain theory~\cite{TurnerW09,LoschP14}, Stone duality~\cite{GLP11,petrisan:phd} and automata theory~\cite{BKL11}.

This paper adds new methodology to this line of research by internalising monoidal categories in nominal sets, where nominal sets are themselves taken to be a monoidal category with respect to the so-called separating tensor.  This allows us to recast categories with a partial monoidal operation as internal categories with a total monoidal operation.

% !TEX root =  calco2019-lmcs.tex

\section{Preliminaries}

From a technical point of view, this paper can be understood as bringing together existing work on string diagrams, or, more specifically, $\PROP$s and nominal sets. In fact, we will be developing the beginnings of a theory of $\PROP$s internal in the category of nominal sets. In this section we review preliminaries on string diagrams and nominal sets.

\hypertarget{sec:props-def}{%
\subsection{String Diagrams, SMTs and PROPs}\label{sec:props-def}}

String diagrams are a 2-(or higher)-dimensional notation for monoidal
categories~\cite{joyal-street:tensor1}. Their algebraic theory can be
formalised by \(\PROP\)s as defined by MacLane~\cite{maclane}. There is
also the weaker notion by Lack~\cite{lack}, see Remark 2.9 of Zanasi~\cite{zanasi} for a discussion.

A \(\label{PROP-def}\PROP\) (\textbf{pro}ducts and \textbf{p}ermutation
category) is a symmetric strict monoidal category, with natural numbers
as objects, where the monoidal tensor \(\oplus\) is addition. Moreover,
\(\PROP\)s, along with strict symmetric monoidal identity-on-objects functors
form the category \(\PROP\). A \(\PROP\) contains
all bijections between numbers as they can be be generated from the
symmetry (twist) \(\sigma:1\oplus 1\to 1\oplus 1\) and from the parallel
composition \(\oplus\) and sequential composition \(;\) (which we write % chktex 40
in diagrammatic order). We denote by \(\sigma_{n,m}\) the canonical
symmetry \(n\oplus m\to m\oplus n\). Functors between \(\PROP\)s
preserve bijections.

\(\PROP\)s can be presented in algebraic form by operations and
equations as \emph{symmetric monoidal theories} (\(\SMT\)s)~\cite{zanasi}.

\begin{figure}[h]
\begin{center}
\begin{tabular}{ c c c }
\includegraphics[page=40, width=12mm]{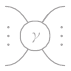} &
\qquad\qquad\includegraphics[page=41, width=12mm]{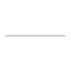}\qquad\qquad{} &
\includegraphics[page=42, width=12mm]{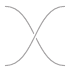} \\
\(\displaystyle\frac{}{ \gen: m \to n \in \Sigma}\) &
\(\displaystyle\frac{}{ id:1 \to 1}\) &
\(\displaystyle\frac{}{\tw : 2 \to 2}\) \\
\end{tabular}
\medskip
\medskip
\medskip

\begin{tabular}{ c c }
\includegraphics[page=39, width=60mm]{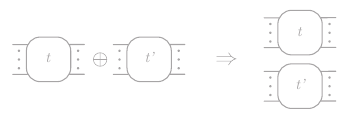} &
\includegraphics[page=38, width=70mm]{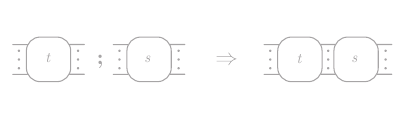} \\
\(\displaystyle\frac{ t:m\to n\quad\quad t':o\to p}{ t \oplus t' :  m+o\to n+p}\) &
\(\displaystyle\frac{ t:m\to n\quad\quad s:n\to o}{ t\hcomp s :  m \to o}\)
\end{tabular}
\end{center}
\caption{SMT Terms (\(\Trm\))}\label{fig:smt-terms}
\end{figure}

An \(\label{SMT-def}\SMT\) \((\Sigma, E)\) has a set \(\Sigma\) of
generators, where each generator \(\gen \in \Sigma\) is given an arity
\(m\) and co-arity \(n\), usually written as \(\gen : m \to n\) and a
set \(E\) of equations, which are pairs of \(\Sigma\)-terms. The set of all $\Sigma$-Terms is denoted by  $\Trm(\Sigma)$.
\(\Sigma\)-terms can be obtained by composing generators in \(\Sigma\)
with the unit \(\id : 1 \to 1\) and symmetry \(\tw : 2 \to 2\), using
either the parallel or sequential composition (see
\autoref{fig:smt-terms}). Equations \(E\) are pairs of \(\Sigma\)-terms
with the same arity and co-arity.

Given an \(\SMT\) \(\langle\Sigma, E\rangle\), we can freely generate a
\(\PROP\), by taking \(\Sigma\)-terms as arrows, modulo the equations
\(\label{SMTeq}\SMTeq\), which are:
\begin{itemize}
\item
  the equations stating that  compositions
  \(;\) and \(\oplus\) form monoids % chktex 40
\item
  the equations of \autoref{fig:symmetric-monoidal-category}
\item
  the equations \(E\)
\end{itemize}
\begin{figure}[h]
\begin{align}
\tag{SMT-sym}%
\label{SMT-sym}
\tw_{1,1} \hcomp \tw_{1,1} = id_2 \qquad{}
\\[1ex]
\tag{SMT-ch}%
\label{SMT-ch}
(s \hcomp t) \oplus (u \hcomp v) = (s \oplus u) \hcomp (t \oplus v)
\\[1ex]
\tag{SMT-nat}%
\label{SMT-nat}
\frac{s : m \to n \qquad t : o \to p}{(s \oplus t) \hcomp \tw_{n,p} = \tw_{m,o} \hcomp (t \oplus s)}
\end{align}
\caption{Equations of symmetric monoidal categories}\label{fig:symmetric-monoidal-category}
\end{figure}

\noindent
\(\PROP\)s have a nice 2-dimensional notation, where sequential
composition is horizontal composition of diagrams, and parallel/tensor
composition is vertical stacking of diagrams (see \autoref{fig:smt-terms}).

\begin{figure}
\includegraphics[page=36, width=\linewidth]{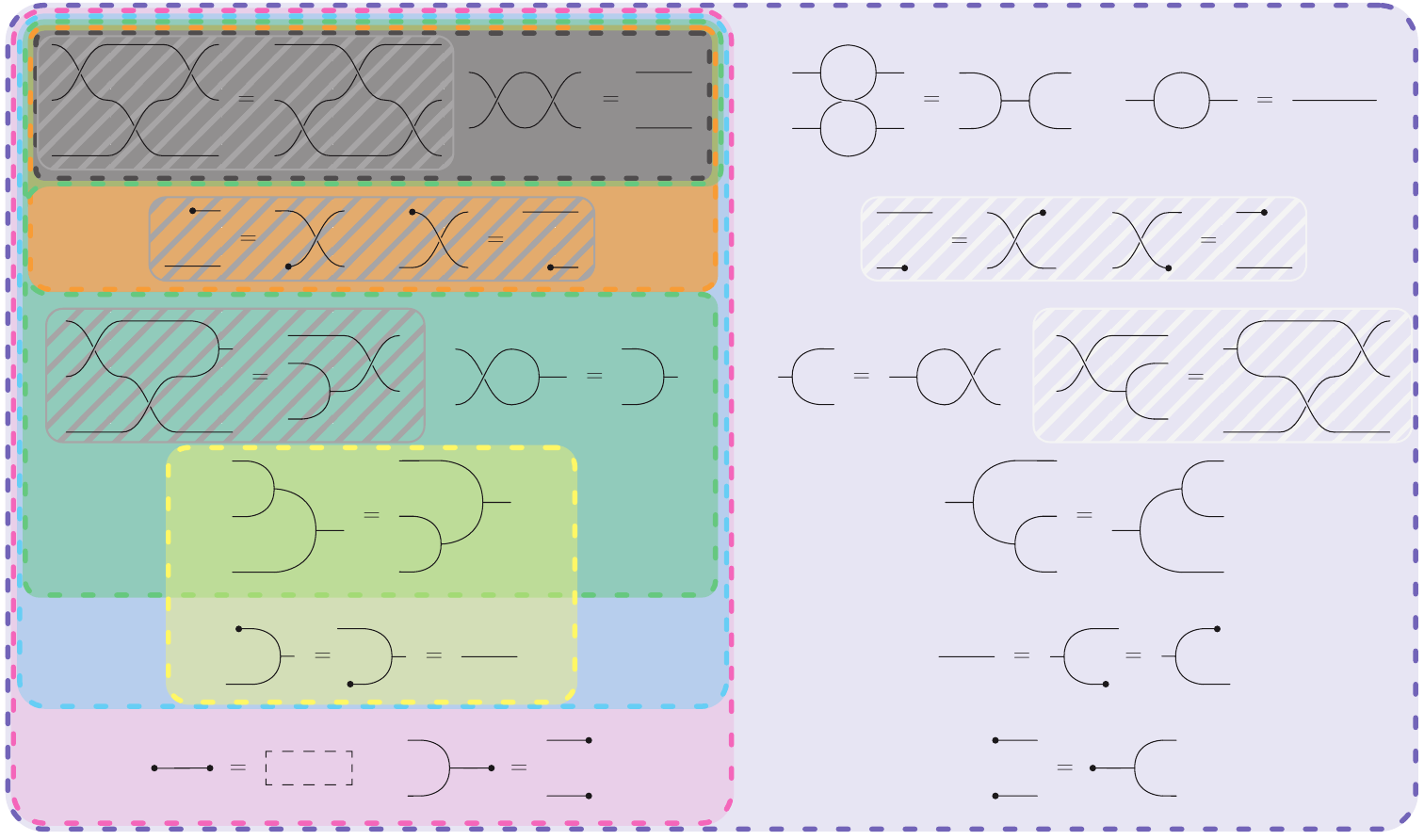}
\caption{Symmetric monoidal theories (compiled from~\cite{lafont})}\label{fig:smt-theories}
\end{figure}

We now present the \(\SMT\)s of \(\bijbox\,\), \(\injbox\,\),
\(\surjbox\,\), \(\funbox\,\), \(\pfunbox\,\), \(\relbox\) and
\(\mmapbox\).\footnote{The theory of \(\mmapbox\) does not include
  equations involving the symmetry \(\tw\) and is in fact presented by a
  so-called \(\PRO\) rather than a \(\PROP\). However, in this paper we
  will only be dealing with theories presented by \(\PROP\)s (the reason
  why this is the case is illustrated in the proof of Proposition~\ref{prop:ORD}).}
The diagram in \autoref{fig:smt-theories} shows the corresponding operations and
equations.

%As explained in the introduction, the aim of this paper is to develop a procedure that will allow us to translate these theories in a systematic way to nominal monoidal theories and to transfer the known completeness results to the nominal setting.

To ease comparison with the corresponding nominal monoidal theories in
\autoref{fig:nmt-theories}, we also added a \stripbox{\strut striped}
background to the equations with wire-crossings, since they are already
implied by the naturality of symmetries~\eqref{SMT-nat}. These are
equations that are part of the definition of a \(\PROP\) in the sense of
MacLane~\cite{maclane} but not in the sense of Lack~\cite{lack}. The
right-hand equation for \(\bijbox\) is~\eqref{SMT-sym} and holds in all
symmetric monoidal theories. We list it here to emphasise the difference
with \autoref{fig:nmt-theories}.

\begin{exa}\label{ex:F}
The \(\SMT\) of $\funbox$ presents the category \(\label{funSkel-def}\FunSkel\) which has natural numbers as objects, all functions as arrows and coproducts as $\oplus$. It is important to note that $\oplus$ is symmetric, but not commutative, with, for example, the symmetry $\sigma_{2,3}$ being  depicted as
\begin{center}
\includegraphics[page=68, width=45mm]{twists_new_thesis}
\end{center}
\end{exa}

We will see later that the category $\nF$ of nominal  finite functions, see Examples~\ref{ex:nF},~\ref{exle:nF-intmoncat} and~\ref{ex:nmt-presentations}, has a commutative tensor.

\hypertarget{nominal-sets}{%
\subsection{Nominal Sets}\label{nominal-sets}}

Let \(\names\) be a countably infinite set of `names' or `atoms'. Let
\(\mathfrak S\) be the group of finite\footnote{A permutation is called
  finite if it is generated by finitely many transpositions.}
permutations \(\names\to\names\,\). An element \(x\in X\) of a group
action \(\mathfrak S\times X\to X\) is \emph{supported} by \(S\subseteq\names\)
if \(\pi\cdot x= x\) for all \(\pi\in\mathfrak S\) and \(x \in S\)
(i.e.~\(\pi\) restricted to \(S\) is the identity). A group action
\(\mathfrak S\times X\to X\) where all elements of \(X\) have finite
support is called a \emph{nominal set}.

We write \(\supp(x)\) for the minimal support of \(x\) and \(\Nom\) for
the category of nominal sets, which has as maps the \emph{equivariant}
functions, that is, those functions that respect the permutation action.
As our running example, we choose the category of simultaneous substitutions:

\begin{exa}\label{ex:nF}
\hypertarget{ex:nF}{}%
We denote by \(\Fun\) the category of finite subsets of \(\names\) as objects and with
all functions as arrows. While \(\Fun\) is a category, it also carries additional
nominal structure. In particular, both the set of objects and the set of
arrows are nominal sets with \(\supp(A)=A\) and \(\supp(f)=A\cup B\) for
\(f:A\to B\). The categories of injections \(\mathsf n\mathbb I\),
surjections \(\mathsf n\mathbb S\), bijections \(\mathsf n\mathbb B\),
partial functions \(\mathsf n\mathbb P\) and relations
\(\mathsf n\mathbb R\) are further examples along the same lines.
\end{exa}

One of the aims of this paper is to exhibit and analyse further structure of this example. For example, all bijections in $\Fun$ can be built from basic functions
\[[a\mapsto b]: \{a\}\to\{b\}\]
and the monoidal operations of sequential and parallel composition as for example in
\[
[a{\mapsto}b]\hcomp[b{\mapsto}c] = [a{\mapsto}c]
\quad\quad
\quad\quad
[a{\mapsto}b] \uplus [c{\mapsto}d] = [a{\mapsto}b, c{\mapsto}d].
\]

We call $\uplus$ the tensor, or the monoidal or vertical or parallel composition. Semantically, the simultaneous substitution on the right-hand side above, will correspond to the function
$f:\{a,c\}\to \{b,d\}$
satisfying $f(a)=b$ and $f(c)=d$.

Importantly, parallel composition of simultaneous substitutions is partial. For example,
$[a{\mapsto}b] \uplus [a{\mapsto}c]$
is undefined, since there is no function $\{a\}\to\{b,c\}$ that maps $a$ simultaneously to both $b$ and $c$.

\begin{rem}
Let us make some remarks about the advantages of a 2-dimensional calculus for simultaneous substitutions over a 1-dimensional calculus one.
A calculus of substitutions can be understood as an algebraic representation of the category $\mathsf n\mathbb F$ of finite subsets of $\names$.  In a 1-dimensional calculus, operations $[a{\mapsto}b]$ have to be indexed by finite sets $S$
\[[a{\mapsto}b]_S:S\cup \{a\}\to S\cup \{b\}\]
for sets $S$ with $a,b\notin S$.
On the other hand, in a  2-dimensional calculus with an explicit operation $\uplus$ for set union, indexing with subsets $S$  is unnecessary. Moreover,  while the swapping
\[\{a,b\}\to\{a,b\}\]
in the 1-dimensional calculus needs an auxiliary name such as $c$ in
$
[a{\mapsto}c]_{\{b\}} \hcomp [b{\mapsto} a]_{\{c\}} \hcomp [c{\mapsto}a]_{\{b\}}
$
it is represented in the 2-dimensional calculus directly by
\[
[a{\mapsto}b] \uplus [b{\mapsto}a]
\]
Finally, while it is possible to write down the equations and rewrite rules for the 1-dimensional calculus, it does not appear as particularly natural. In particular, only in the 2-dimensional calculus, will the swapping
%substitution $[a{\mapsto}c]_{\{b\}} \hcomp [b{\mapsto} a]_{\{c\}} \hcomp [c{\mapsto}a]_{\{b\}}$
have a simple normal form such as $[a{\mapsto}b] \uplus [b{\mapsto}a]$ (unique up to commutativity of $\uplus$).
\end{rem}

% !TEX root =  calco2019-lmcs.tex

\hypertarget{sec:internal-monoidal}{%
\section{Internal monoidal categories}\label{sec:internal-monoidal}}

We introduce the, to our knowledge, novel notion of an internal monoidal category. Given a
symmetric monoidal category \((\mathcal V,I,\otimes)\) with finite
limits, we are interested in categories \(\mathbb C\), internal in
\(\mathcal V\), that carry a monoidal structure not of type
\(\mathbb C\times \mathbb C\to \mathbb C\) but of type
\(\mathbb C\lotimes \mathbb C\to \mathbb C\), where $\lotimes$ is a certain lifting, defined below, of $\otimes$ to categories internal in $\mathcal V$.

Before going into the technicalities, let us resume Example~\ref{ex:nF} from the previous section in order to explain why we want to work with the lifted tensor product $\lotimes$. First, we give the example of the monoidal category in which we will internalise.

\begin{exa} The symmetric monoidal (closed) category
 \(\label{Nom-cat-def}(\Nom,1,\ast )\) of nominal sets with the separated
 product \(\ast \) is defined as follows~\cite{pitts}. \(1\) is the
 terminal object, i.e.~a singleton with empty support. The separated
 product of two nominal sets is defined as
 \(A \ast B = \{(a,b)\in A\times B \mid \supp(a)\cap\supp(b)=\emptyset\}\). $\Nom$ is also a symmetric monoidal (closed) category wrt the cartesian product. The two monoidal structures are related by injections $j_{A,B}:A\ast B\to A\times B$ natural in $A$ and $B$.
\end{exa}

Next, let us go back to the category $\Fun$ of Example~\ref{ex:nF}, which is the category we want to internalise in $(\Nom,1,\ast )$. Recall that while parallel composition in the category $\mathbb F$ of Example~\ref{ex:F} is the coproduct
\[\oplus : \mathbb F\times \mathbb F\to\mathbb F\]
we have seen in Example~\ref{ex:nF} that the parallel composition in $\Fun$ is a partial operation
\[\uplus:\Fun\times\Fun\to\Fun.\]
One way to proceed would be to develop a theory of partial monoidal categories. But in our case, it seems more elegant to notice that $\uplus$ is a total operation
\[\uplus:\Fun\ast\Fun\to\Fun\]
since the separated product $\ast$ accounts for the constraint that $f\uplus g$ is defined iff the domain and codomain of $f$ and $g$ are disjoint.
The purpose of this section is to
define the notion of internal monoidal category and to show that
\((\Fun,\emptyset,\uplus)\) is a  monoidal category internal in
\((\Nom,1,\ast)\) with \(\uplus\) of type
\[\uplus:\Fun\last\Fun\to\Fun.\]
where $\last$ is the lifting of $\ast$ to categories internal in $\Nom$.

The task, then, of this section is to extend
\(\ast:\Nom\times\Nom\to\Nom\) to
\[\last:\Cat(\Nom)\times\Cat(\Nom)\to\Cat(\Nom)\] where we denote by
\(\Cat(\Nom)\), the category of small internal categories%
\footnote{The
 necessary notation from internal categories is reviewed in Appendix~\ref{app:internalCats}.\\} in \(\Nom\). See also Remark~\ref{rem:intmoncat-summary} for a summary of notation. At the end of the section, we will have justified our main example:
\begin{exa}\label{exle:nF-intmoncat}
 Following on from the previous example, the category \(\Fun\) is a monoidal category internal in \((\Nom,1,\ast )\) with monoidal
 operation given by \(A \uplus B = A \cup B\) if \(A,B\) are disjoint and
 \(f\uplus f' = f \cup f'\) if \(A,A'\) and \(B,B'\) are disjoint where
 \(f:A \to B\) and \(f':A' \to B'\). The unit is the empty set $\emptyset$.
\end{exa}

Since \(\otimes\) does not need to preserve finite limits, we cannot expect that
defining \((\mathbb C\lotimes\mathbb C)_0=\mathbb C_0\otimes \mathbb C_0\) and
\((\mathbb C\lotimes\mathbb C)_1=\mathbb C_1\otimes \mathbb C_1\) results in
\(\mathbb C\lotimes\mathbb C\) being an internal category. The purpose of the next example is to show what
goes wrong in a concrete instance.

\begin{exa}
Following on from the previous example, given \((\Nom,1,\ast )\), we define, just for the purposes of this example,
a binary operation \(\Fun\last\Fun\) as
\((\Fun\last\Fun)_0=\Fun_0\ast \Fun_0\) and
\((\Fun\last\Fun)_1=\Fun_1\ast \Fun_1\). Then \(\Fun\last\Fun\) cannot be
equipped with the structure of an internal category. Indeed, assume for
a contradiction that there was an appropriate pullback
\((\Fun\last \Fun)_2\) and arrow \(\comp\) such that the two diagrams
commute: \[\xymatrix@C=20ex@R=3ex{
\ \ \,(\Fun\last\Fun)_2 \
\ar[0,1]|-{\ \comp\ }
\ar[dd]_{\pi_1}^{\pi_2}
&
{\ \Fun_1\ast\Fun_1\ }
%\ar@<1ex>[0,1]^-{\dom\times \dom}
%\ar@<-1ex>[0,1]_-{\cod\times \cod}
\ar[dd]_{\dom}^{\cod}
%&
%\ \Fun_0\ast \Fun_0
%\ar[dd]_{\uplus_0}
%
\\
&
\\
\ \ \Fun_1\ast\Fun_1\ \
\ar[0,1]^{\ \dom\ }_{\cod}
&
{\ \ \Fun_0\ast\Fun_0 \ \ }
%\ar@<1ex>[0,1]^-{\dom}
%\ar@<-1ex>[0,1]_-{\cod}
%&
%\ \Fun_0
}\]

Let \(\delta_{xy}:\{x\}\to\{y\}\) be the unique function in \(\Fun\) of
type \(\{x\}\to\{y\}\).
Then \[((\delta_{ac},\delta_{bd}), (\delta_{cb},\delta_{da})),\]
which can be depicted as \[\xymatrix@R=0.5ex{
\{a\} \ar[r]^{\delta_{ac}} & \{c\} \ar[r]^{\delta_{cb}} & \{b\}\\
\{b\} \ar[r]_{\delta_{bd}} & \{d\} \ar[r]_{\delta_{da}} & \{a\}
}\] is in the pullback \((\Fun\last\Fun)_2\), but there is no \(\comp\)
such that the two squares above commute, since
\(\comp((\delta_{ac},\delta_{bd}), (\delta_{cb},\delta_{da}))\) would have to be
\((\delta_{ab},\delta_{ba})\). But since \(\delta_{ab}\) and \(\delta_{ba}\) do not have
disjoint support (since \(\supp(\delta_{ab}) = \supp(\delta_{ba}) = \{a,b\}\)),
this set cannot be in \(\Fun_1\ast \Fun_1\). \qed
\end{exa}

In the example, the attempt to define a tensor on the  category $\mathbb C=\nF$ internal in $(\mathcal V,I,\otimes)=(\Nom,1\ast)$ via $(\mathbb C\otimes \mathbb C)_1 =
\mathbb C_1\otimes \mathbb C_1$ fails. To ask for pairs of arrows in $(\mathbb C\otimes \mathbb C)_1$ to have disjoint support is too much. Instead we should be looking for a general categorical definition that restricts to those pairs of arrows in $(\mathbb C\otimes \mathbb C)_1$ that have disjoint domains and disjoint codomains.

The solution to the problem consists in assuming that the given
symmetric monoidal category with finite limits
\((\mathcal V,1,\otimes)\) is semi-cartesian (aka affine), that is, the
unit \(1\) is the terminal object. In such a category there are
canonical arrows natural in \(A\) and \(B\) (dropping the subscripts of $j_{A,B}$)
\[j:A\otimes B\to A\times B\] and we can use them to define arrows
\(j_1:(\mathbb C\lotimes \mathbb C)_1\to\mathbb C_1\times \mathbb C_1\)
that give us the right notion of tensor on arrows. From our example
\(\Fun\) above, we know that we want arrows
 \((f,g)\) to be in
\((\mathbb C\lotimes \mathbb C)_1\) if \(\dom(f)\cap\dom(g)=\emptyset\)
and \(\cod(f)\cap\cod(g)=\emptyset\). We now turn this into a category
theoretic definition, which is in fact an instance of the general and
well-known construction of pulling back an internal category
\(\mathbb C\) along an arrow \(j:X\to\mathbb C_0\). This construction
yields an internal category \(\mathbb X\) with \(\mathbb X_0=X\) and
\(\mathbb X_1\) the pullback of
\(\langle\dom_{\mathbb C}, \cod_{\mathbb C}\rangle\) along
\(j\times j\), or, equivalently, the limit in the following diagram

\begin{center}
\begin{tikzpicture}
 \matrix (m) [matrix of math nodes, row sep=0.2em, row 2/.style= {yshift=-4em}, column sep=1.6em, color=black]{
                 & |[color=cyan]|\mathbb X_1 &                &                 & \mathbb C_1 & \\
                 &                & \mathbb X_0 &                 &                & \mathbb C_0 \\
  \mathbb X_0 &                &                & \mathbb C_0  &                & \\};
 \path[-stealth, color=black]
  (m-1-2) edge [color=cyan] node [above] {\scriptsize $j_1$} (m-1-5)
      edge [color=cyan, densely dotted] node [right] {\scriptsize $\cod_{\mathbb X}$} (m-2-3)
      edge [color=cyan] node [left] {\scriptsize $\dom_{\mathbb X}$} (m-3-1)
  (m-2-3) edge [densely dotted] node [above, xshift=2.5em] {\scriptsize $j$} (m-2-6)
  (m-3-1) edge node [below] {\scriptsize $j$} (m-3-4)
  (m-1-5) edge [densely dotted] node [right] {\scriptsize $\ \cod_{\mathbb C}$} (m-2-6)
      edge [-,line width=6pt,draw=white] (m-3-4) edge node [left,xshift=1.2em, yshift=1.5em] {\scriptsize $\dom_{\mathbb C}\ \ \ $} (m-3-4);
\end{tikzpicture}
\end{center}

which we abbreviate to \begin{equation}%
\label{equ:j1}
\vcenter{
\xymatrix@C=9ex{
\mathbb X_1 \ar[rr]^{j_1}
\ar@<-1ex>[d]_{\dom_{\mathbb X}}
\ar@<1ex>[d]^{\cod_{\mathbb X}}
&& \mathbb C_1
\ar@<-1ex>[d]_{\dom_{\mathbb C}}
\ar@<1ex>[d]^{\cod_{\mathbb C}}
\\
\mathbb X_0 \ar[rr]_j
&& \mathbb C_0
}}
\end{equation}

Next we define \(i:\mathbb X_0 \to \mathbb X_1\) as the arrow into the
limit \(\mathbb X_1\) given by

\begin{equation}
\vcenter{
\xymatrix@C=9ex{
\mathbb X_0
\ar@/^1.3pc/[rrrd]^{i_{\mathbb C}\circ j}
\ar@/^-1.3pc/[rdd]|-(0.45){\rule[-1mm]{0mm}{3mm}\ \id}
\ar@/^-1.3pc/@<-1ex>[rdd]|-(0.55){\rule[-1mm]{0mm}{3.5mm}\id\ }
\ar@{..>}[rd]|-{\rule[-2mm]{0mm}{5mm}\ i_{\mathbb X}\ }
&&&
\\
&\mathbb X_1 \ar[rr]^{j_1}
\ar@<-1ex>[d]_{\dom_{\mathbb X}}
\ar@<1ex>[d]^{\cod_{\mathbb X}}
&& \mathbb C_1
\ar@<-1ex>[d]_{\dom_{\mathbb C}}
\ar@<1ex>[d]^{\cod_{\mathbb C}}
\\
&{\mathbb X_0} \ar[rr]_j
&& \mathbb C_0
}}%
\label{equ:idef}
\end{equation}

from which one reads off
\[\dom_{\mathbb X}\circ i_{\mathbb X} = \id_{\mathbb X_0} = \cod_{\mathbb X}\circ i_{\mathbb X}\]
Next, \(\mathbb X_2\) is the pullback \[\xymatrix{
&\mathbb X_2\ar[dl]_{\pi_{\mathbb X1}}\ar[dr]^{\pi_{\mathbb X2}}&
\\
\mathbb X_1\ar[dr]_{\cod_{\mathbb X}}
&&
\mathbb X_1 \ar[dl]^{\dom_{\mathbb X}}
\\
&\mathbb X_0&
}\] Recalling the definition of \(j_1\) from~\eqref{equ:j1}, there is
also a corresponding \(j_2:\mathbb X_2\to\mathbb C_2\) due to the fact
that the product of pullbacks is a pullback of products.
\begin{equation}\label{equ:j2def}
\vcenter{
\xymatrix@C=6ex{
&\mathbb X_2
 \ar[dl]|-(0.35){\rule[-2mm]{0mm}{5mm} \pi_{\mathbb X1}}
 \ar[dr]|-(0.35){\rule[-2mm]{0mm}{5mm} \pi_{\mathbb X2}}
 \ar[rrr]^{j_2}&
&
&\mathbb C_2
 \ar[dl]|-(0.35){\rule[-2mm]{0mm}{5mm} \pi_{\mathbb X1}}
 \ar[dr]|-(0.35){\rule[-2mm]{0mm}{5mm} \pi_{\mathbb C2}}&
\\
\mathbb X_1
 \ar[dr]_{\cod_{\mathbb X}}
 \ar@/^1.2pc/@{..>}[rrr]|-(.3){\ j_1\ }
&&
\mathbb X_1
 \ar[dl]^{\dom_{\mathbb X}}
 \ar@/_1.2pc/@{..>}[rrr]|-(.7){\ j_1\ }
&
\mathbb C_1
 \ar[dr]_{\cod_{\mathbb C}}
&&
\mathbb C_1
 \ar[dl]^{\dom_{\mathbb C}}
\\
&\mathbb X_0
 \ar[rrr]_{j}&
&
&\mathbb C_0&
}}
\end{equation}

Recall the definition of the limit \(\mathbb X_1\) from~\eqref{equ:j1}.
Then \(\comp_{\mathbb X}:\mathbb X_2 \to \mathbb X_1\) is the arrow into
\(\mathbb X_1\) \begin{equation}\label{equ:comp}
\vcenter{
\xymatrix@C=11ex{
\mathbb X_2
\ar@/^/[rrrd]^{\comp_{\mathbb C}\circ j_2}
\ar@/^-2pc/[rdd]^(0.4){\cod_{\mathbb X}\circ\pi_{\mathbb X2}}
\ar@/^-2pc/@<-1ex>[rdd]_(0.4){\dom_{\mathbb X}\circ\pi_{\mathbb X1}}
\ar@{..>}[rd]|-{\rule[-2mm]{0mm}{5mm} \comp_{\mathbb X}}
&&&
\\
&\mathbb X_1 \ar[rr]^{j_1}
\ar@<-1ex>[d]_{\dom_{\mathbb X}}
\ar@<1ex>[d]^{\cod_{\mathbb X}}
&& \mathbb C_1
\ar@<-1ex>[d]_{\dom_{\mathbb C}}
\ar@<1ex>[d]^{\cod_{\mathbb C}}
\\
&{\mathbb X_0} \ar[rr]_j
&& \mathbb C_0
}}
\end{equation}

from which one reads off
\[\dom_{\mathbb X}\circ \comp_{\mathbb X}=\dom_{\mathbb X}\circ\pi_{\mathbb X1}\ \quad \
\cod_{\mathbb X}\circ \comp_{\mathbb X}=\cod_{\mathbb X}\circ\pi_{\mathbb X2}\ \quad \
j_1 \circ \comp_{\mathbb X}=\comp_{\mathbb C}\circ j_2\]

and the remaining equations
\(\comp_{\mathbb X}\circ \langle i_{\mathbb X}\circ\dom_{\mathbb X},\id_{\mathbb X_1} \rangle = \id_{\mathbb X_1} = \comp_{\mathbb X}\circ \langle\id_{\mathbb X_1},i_{\mathbb X}\circ\cod_{\mathbb X} \rangle\)
are also not difficult to prove.

Finally, in analogy with the definition of \(j_2\) in~\eqref{equ:j2def},
\(j_3\) is defined as the unique arrow into the pullback
\(\mathbb C_3\), where \(\mathbb X_3\) is defined in the expected way:

\begin{equation}\label{equ:j3def}
\vcenter{
\xymatrix@C=6ex{
&\mathbb X_3
 \ar[dl]|-(0.35){\rule[-2mm]{0mm}{5mm} \leftt_{\mathbb X}}
 \ar[dr]|-(0.35){\rule[-2mm]{0mm}{5mm} \rightt_{\mathbb X}}
 \ar[rrr]^{j_3}&
&
&\mathbb C_3
 \ar[dl]|-(0.35){\rule[-2mm]{0mm}{5mm} \leftt_{\mathbb C}}
 \ar[dr]|-(0.35){\rule[-2mm]{0mm}{5mm} \rightt_{\mathbb C}}&
\\
\mathbb X_2
 \ar[dr]_{\pi_{\mathbb X2}}
 \ar@/^1.2pc/@{..>}[rrr]|-(.3){\ j_2\ }
&&
\mathbb X_2
 \ar[dl]^{\pi_{\mathbb X1}}
 \ar@/_1.2pc/@{..>}[rrr]|-(.7){\ j_2\ }
&
\mathbb C_2
 \ar[dr]_{\pi_{\mathbb C2}}
&&
\mathbb C_2
 \ar[dl]^{\pi_{\mathbb C1}}
\\
&\mathbb X_1
 \ar[rrr]_{j_1}&
&
&\mathbb C_1&
}}
\end{equation}

This ends the definition of \(\mathbb X\).  Since pulling back internal categories is a well known construction, we skip the verification that  \(\mathbb X\) is an internal category. The reader interested in the technical details may find them in the thesis~\cite{balco:phd} of the first author. We summarise what we have done so far in
\begin{prop}%
\label{prop:pullbackofinternalcategories}%
Given an internal category \(\mathbb C\) and an arrow
\(j:X\to\mathbb C_0\) there is an internal category \(\mathbb X\) and an
internal functor \(\mathbb j:\mathbb X\to\mathbb C\) such that
\(\mathbb X_0=X\) and \(\mathbb j_0=j\).
\end{prop}
To continue our development, we specialise to $\mathbb j: \mathbb C\otimes \mathbb C\to \mathbb C\times \mathbb C$ available in all semi-cartesian monoidal categories.
To make progress, we need to show that $\mathbb C\otimes \mathbb C$ extends to a functor $\lotimes:\Cat(\mathcal V)\times \Cat(\mathcal V)\to \Cat(\mathcal V)$ and that the $\mathbb j: \mathbb C\lotimes \mathbb C\to \mathbb C\times \mathbb C$ are natural in $\mathbb C$.
As usual, the easiest way of proving functoriality and naturality is to exhibit the construction in question, in our case $\lotimes$, as a universal one.
Thus, the next step is to exhibit the \(\mathbb j:\mathbb X\to\mathbb C\) from the proposition as a so-called cartesian lifting, a notion from the theory of fibrations~\cite{jacobs,streicher}.

\begin{defi}[Fibration]
If \(P:\mathcal W\to\mathcal V\) is a functor, then
\(\mathbb j : \mathbb X\to \mathbb Y\) is a \emph{cartesian lifting} of
\(j:X\to P\mathbb Y\) if for all
\(\mathbb k:\mathbb W\to \mathbb Y\) and all
\(h:P\mathbb W\to X\) with \(P\mathbb k=j\circ h\) there is a
unique \(\mathbb h:\mathbb W\to\mathbb X\) such that
\(\mathbb j\circ\mathbb h=\mathbb k\) and \(P\mathbb h=h\).
\[\xymatrix@C=9ex{
\mathbb W
\ar[rrdd]^{\mathbb k}
\ar@{..>}[rdd]_{!\mathbb h}
&&
\\
\\
&\mathbb X
\ar[r]_{\mathbb j}
&\mathbb Y
}\qquad\xymatrix@C=9ex{
P\mathbb W
\ar[rrdd]^{P\mathbb k = j \circ h}
\ar[rdd]_{h}
&&
\\
\\
&X
\ar[r]_{j}
&P\mathbb Y
\\
}\] Moreover, \(P:\mathcal W\to\mathcal V\) is called a (Grothendieck)
\emph{fibration} if all \(j:X\to P\mathbb Y\) have a cartesian
lifting for all \(\mathbb Y\) in \(\mathcal W\). If
\(P:\mathcal W\to\mathcal V\) is a fibration, the subcategory of
\(\mathcal W\,\) that has as arrows the arrows \(\mathbb f\) such that
\(P\mathbb f=\id_Y\) is called the \emph{fibre} over \(Y\).
\end{defi}

The next lemma is a strengthening of Proposition~\ref{prop:pullbackofinternalcategories}.

\begin{lem}%
\label{lem:catvv-is-a-fibration}%
Let \(\mathcal V\) be a category with finite limits. The forgetful
functor \(\Cat(\mathcal V)\to\mathcal V\) is a fibration.

\begin{proof}
We have already shown how to lift \(j:X\to\mathbb C_0\) to
\(\mathbb j:\mathbb X\to\mathbb C\). One can show that this is a
cartesian lifting by drawing out the appropriate diagram. Namely, we
have the forgetful functor \((-)_0:\Cat(\mathcal V)\to\mathcal V\),
which sends an internal category to its ``object of objects'',
internal categories \(\mathbb X\), \(\mathbb Y\) and an internal functor
\(\mathbb j\) between them. Given another internal category
\(\mathbb W\) and an internal functor
\({\color{magenta}\mathbb k} : \mathbb W \to\mathbb Y\) and an arrow
\({\color{cyan}h} : \mathbb W_0 \to \mathbb Y_0\), s.t.
\({\color{magenta}\mathbb k_0} = \mathbb j_0 \circ {\color{cyan}h}\), we
show there is a unique \(\color{cyan} \mathbb h\), s.t.
\({\color{magenta}\mathbb k} = \mathbb j \circ {\color{cyan}\mathbb h}\).
This essentially means we need to fill in the following diagram, such
that all sub-diagrams commute:

\[\vcenter{
\xymatrix@C=12ex{
\mathbb W_2
 \ar@/^1.5pc/@[magenta][rr]|-{\color{magenta}\ \mathbb k_2\ }
 \ar@/^-1.2pc/[d]_{\pi_1}
 \ar@/^1.2pc/[d]^{\pi_2}
 \ar[d]|-{\rule[-1mm]{0mm}{3mm} \comp}
&
\mathbb X_2
 \ar[r]^{\mathbb j_2}
 \ar@/^-1.2pc/[d]_{\pi_1}
 \ar@/^1.2pc/[d]^{\pi_2}
 \ar[d]|-{\rule[-1mm]{0mm}{3mm} \comp}
&
\mathbb Y_2
 \ar@/^-1.2pc/[d]_{\pi_1}
 \ar@/^1.2pc/[d]^{\pi_2}
 \ar[d]|-{\rule[-1mm]{0mm}{3mm} \comp}
\\
\mathbb W_1
 \ar@/^-1.2pc/@[magenta][rr]|-(.80){\color{magenta}\ \mathbb k_1\ }
 \ar@<-1ex>[d]_{\dom_{\mathbb W}}
 \ar@<1ex>[d]^{\cod_{\mathbb W}}
&
\mathbb X_1
 \ar[r]^{\mathbb j_1}
 \ar@<-1ex>[d]_{\dom_{\mathbb X}}
 \ar@<1ex>[d]^{\cod_{\mathbb X}}
&
\mathbb Y_1
 \ar@<-1ex>[d]_{\dom_{\mathbb Y}}
 \ar@<1ex>[d]^{\cod_{\mathbb Y}}
\\
\mathbb W_0
 \ar@/^-1.5pc/@[magenta][rr]|-{\color{magenta}\ \mathbb k_0\ }
 \ar@[cyan][r]_{\color{cyan} h}
 \ar[u]|-{\rule[-2mm]{0mm}{0mm} i_{\mathbb W}}
&
\mathbb X_0
 \ar[r]_{\mathbb j_0}
 \ar[u]|-{\rule[-2mm]{0mm}{0mm} i_{\mathbb X}}
&
\mathbb Y_0
 \ar[u]|-{\rule[-2mm]{0mm}{0mm} i_{\mathbb Y}}
}}\]

Since our category has all finite limits, we can define
\(\color{cyan}\mathbb h_1\) as an arrow into the limit \(\mathbb X_1\):

\[\xymatrix@C=9ex{
\mathbb W_1
\ar@/^1.3pc/@[magenta][rrrd]^{\color{magenta} \mathbb k_1}
\ar@/^-1.3pc/[rdd]^(0.45){\!\cod_{\mathbb W} \circ\,\color{cyan} h}
\ar@/^-1.3pc/@<-1ex>[rdd]_{\dom_{\mathbb W} \circ\,\color{cyan} h}
\ar@{..>}@[cyan][rd]|-{\rule[-2mm]{0mm}{5mm}\ \color{cyan}\mathbb h_1\ }
&&&
\\
&\mathbb X_1 \ar[rr]^{\mathbb j_1}
\ar@<-1ex>[d]_{\dom_{\mathbb X}}
\ar@<1ex>[d]^{\cod_{\mathbb X}}
&& \mathbb Y_1
\ar@<-1ex>[d]_{\dom_{\mathbb Y}}
\ar@<1ex>[d]^{\cod_{\mathbb Y}}
\\
&{\mathbb X_0} \ar[rr]_{\mathbb j_0}
&& \mathbb Y_0
}\]

We obtain \(\color{cyan}\mathbb h_2\) in a similar fashion, thus getting
a unique \(\color{cyan}\mathbb h = (\mathbb h_2, \mathbb h_1, h)\), for
which we have
\({\color{magenta}\mathbb k} = \mathbb j \circ {\color{cyan}\mathbb h}\).
\end{proof}
\end{lem}

%Instantiating \autoref{lem:catvv-is-a-fibration} with
%\(\mathbb C\times \mathbb D \in \Cat(\mathcal V)\) for \(\mathbb Y\) and
%\(j:\mathbb C_0\otimes \mathbb D_0 \to \mathbb C_0\times \mathbb D_0\)
%for \(j:\mathbb X_0\to\mathbb Y_0\), gives us the desired result from
%the beginning of this section, namely that the tensor \(\otimes \) in a
%symmetric monoidal category \((\mathcal V,1,\otimes)\) can be lifted to
%a tensor
%\(\lotimes : \Cat(\mathcal V) \times \Cat(\mathcal V)\to \Cat(\mathcal V)\):
%
%
%\begin{cor}
%The arrow
%\(j:\mathbb C_0\otimes \mathbb D_0 \to \mathbb C_0\times \mathbb D_0\)
%lifts to a morphism of internal categories
%\(\mathbb j: \mathbb C\lotimes \mathbb D \to \mathbb C\times \mathbb D\).
%Moreover, \(\mathbb j\) is the cartesian lifting of \(j\).
%\end{cor}

%Instantiating \autoref{lem:catvv-is-a-fibration} with
%\(\mathbb C\times \mathbb D \in \Cat(\mathcal V)\) for \(\mathbb Y\) and
%\(j:\mathbb C_0\otimes \mathbb D_0 \to \mathbb C_0\times \mathbb D_0\)
%for \(j:\mathbb X_0\to\mathbb Y_0\), gives us the desired result from
%the beginning of this section, namely that the tensor \(\otimes \) in a
%symmetric monoidal category \((\mathcal V,1,\otimes)\) can be lifted to
%a tensor
%\(\lotimes : \Cat(\mathcal V) \times \Cat(\mathcal V)\to \Cat(\mathcal V)\):

While both Proposition~\ref{prop:pullbackofinternalcategories} and Lemma~\ref{lem:catvv-is-a-fibration} allow us to conclude that
\(\otimes\) on $\mathcal V$
can be lifted to an operation \(\lotimes\) on $\Cat(\mathcal V)$, we rely on the universal property of Lemma~\ref{lem:catvv-is-a-fibration} to argue that $\lotimes$ is functorial and that $\mathbb j$ is natural.
To show that $\lotimes$ is functorial we use that
\(\otimes:\mathcal V\times \mathcal V\to\mathcal V\) is functorial and
that
\(j:\mathbb C_0\otimes \mathbb D_0 \to \mathbb C_0\times \mathbb D_0\)
is natural in \(\mathbb C\) and \(\mathbb D\). In order to lift such
natural transformations, which are arrows in the functor category
\(\mathcal V^{\Cat(\mathcal V)\times\Cat(\mathcal V)}\), we use

\begin{lem}\label{lem:fib-lifted}%
If \(P:\mathcal W\to\mathcal V\) is a fibration and \(\mathcal A\) is a
category, then
\(P^{\mathcal A}:\mathcal W^{\mathcal A}\to\mathcal V^{\mathcal A}\)
is a fibration.

\begin{proof}
\(P^{\mathcal A}\) is defined via post-composition with \(P\), that is, \(P^{\mathcal A}(\mathbb G) = P \circ \mathbb G=P\mathbb G\)
and \(P^{\mathcal A}(\eta : \mathbb G \to \mathbb H) = P\eta \).
To show that \(P^{\mathcal A}\) is a fibration, i.e.~that all \(j :G \to P\mathbb H\) have a \emph{cartesian
lifting} \(\mathbb j : \mathbb G \to \mathbb H\), we lift \(j\) point-wise, using the fact
that for all \(j_A : GA \to P\mathbb HA\) we have
\(\mathbb j_A : \mathbb G A \to \mathbb HA\) due to \(P\) being a fibration. It
remains to check that \(\mathbb j\) is a \emph{cartesian
lifting}, that is, given natural transformations \(\mathbb k : \mathbb F \to \mathbb H\)
and \(h : P\mathbb F \to G\), such that
\(P\mathbb k = P\mathbb j \circ h\), there is
a unique \(\mathbb h\), s.t.\ the following diagrams commute
\[
\xymatrix@C=9ex{
\mathbb F
\ar@{=|>}@/^1.3pc/[rrdd]^{\mathbb k}
\ar@{:|>}[rdd]_{!\mathbb h}
&&
\\
\\
&\mathbb G
\ar@{=|>}[r]_{\mathbb j}
&\mathbb H}\qquad
\xymatrix@C=9ex{
P\mathbb F
\ar@{=|>}@/^1.3pc/[rrdd]^{\ \ P\mathbb k \,=\, P\mathbb j \circ h}
\ar@{=|>}[rdd]_{P\mathbb h = h}
&&
\\
\\
&G
\ar@{=|>}[r]_{j \,=\, P\mathbb j}
&P \mathbb H
\\
}\]

Since \(\mathbb k\), \(\mathbb j\) and \(h\) are natural
transformations we have for all \(f : A \to B\)
\begin{equation*}
\xymatrix@C=9ex{
{\color{magenta} \mathbb FA}
\ar[r]^{\mathbb Ff}
\ar@[magenta][d]^{\color{magenta} \mathbb k_A}
&
{\color{cyan} \mathbb FB}
\ar@[cyan][d]_{\color{cyan} \mathbb k_B}
\\
{\color{magenta} \mathbb HA}
\ar[r]^{\mathbb Hf}
&
{\color{cyan} \mathbb HB}
\\
{\color{magenta} \mathbb GA}
\ar[r]_{\mathbb Gf}
\ar@[magenta][u]_{\color{magenta} \mathbb j_A}
&
{\color{cyan} \mathbb GB}
\ar@[cyan][u]^{\color{cyan} \mathbb j_B}
}
\quad\quad
\quad\quad
%%%%%%%%%
\xymatrix{
{\color{magenta} P\mathbb FA}
\ar[r]^{P\mathbb Ff}
\ar@[magenta][d]^{\color{magenta} P\mathbb k_A}
\ar@[magenta]@/^-1.3pc/[dd]_{\color{magenta} h_A}
&
{\color{cyan} PFB}
\ar@[cyan][d]_{\color{cyan} P\mathbb k_B}
\ar@[cyan]@/^1.3pc/[dd]^{\color{cyan} h_B}
\\
{\color{magenta}P\mathbb HA}
\ar[r]^{P\mathbb Hf}
&
{\color{cyan}P\mathbb HB}
\\
{\color{magenta} P\mathbb GA}
\ar[r]_{P\mathbb Gf}
\ar@[cyan]@[magenta][u]_{\color{magenta} P\mathbb j_A}
&
{\color{cyan} P\mathbb GB}
\ar@[cyan][u]^{\color{cyan} P\mathbb j_B}
}
\end{equation*}

As \(P\) is a fibration, we obtain unique \(\mathbb h_A\) and
\(\mathbb h_B\) for the diagram on the left above, s.t.
\(P\mathbb h_A = h_A\) and \(P\mathbb h_B = h_B\), thus obtaining a
unique natural transformation \(\mathbb h\), for which
\(\mathbb k = \mathbb h \circ \mathbb j\).
\end{proof}
\end{lem}

Instantiating the lemma with
\(P=(-)_0:\Cat(\mathcal V)\to\mathcal V\) and
\(\mathcal A=\Cat(\mathcal V)\times\Cat(\mathcal V)\), we obtain as
a corollary that lifting the tensor
\(\otimes: \mathcal V\times \mathcal V\to \mathcal V\) to
\(\lotimes: \Cat(\mathcal V)\times \Cat(\mathcal V)\to \Cat(\mathcal V)\)
is functorial:

\begin{thm}
Let \((\mathcal V,1,\otimes)\) be a (symmetric) monoidal category with
finite limits in which the monoidal unit is the terminal object. Let
\(P=(-)_0:\Cat(\mathcal V)\to\mathcal V\)  be the forgetful functor from
categories internal in \(\mathcal V\). Then the canonical arrow
\(j:\mathbb C_0\otimes \mathbb D_0 \to \mathbb C_0\times \mathbb D_0\)
lifts uniquely to a natural transformation
\(\mathbb j:\mathbb C\lotimes \mathbb D \to \mathbb C\times \mathbb D\) and the terminal object $1\in\mathcal V$ lifts uniquely to a monoidal unit $\mathbb I$ of $\lotimes$.
Moreover, \((\Cat(\mathcal V), \mathbb I, \lotimes)\) inherits from
\((\mathcal V, 1, \otimes)\) the structure of a (symmetric) monoidal
category with finite limits in which the monoidal unit is the terminal
object.

\begin{proof}
We continue to use the notation of the proof of Lemma~\ref{lem:fib-lifted}.
Let
{\small
\begin{align*}
  &\mathbb H:\Cat(\mathcal V) \times \Cat(\mathcal V)\to\Cat(\mathcal V) \qquad
  &G&:\Cat(\mathcal V) \times \Cat(\mathcal V) \to \mathcal V \qquad
  &\mathbb G&:\Cat(\mathcal V) \times \Cat(\mathcal V)\to\Cat(\mathcal V) \\
  &\mathbb H\,(\mathbb A, \mathbb B) = \mathbb A\times \mathbb B
  &G& (\mathbb A, \mathbb B) = P\mathbb A\otimes P\mathbb B
  &\mathbb G& (\mathbb A, \mathbb B) = \mathbb A\lotimes \mathbb B
\end{align*}
}%
and \(j:G \to P\mathbb H\) be the associated natural
transformation. We have by definition that \(G = P\mathbb G\),
that is,
\(\mathbb A_0 \otimes \mathbb B_0 = (\mathbb A \lotimes \mathbb B)_0\).
Therefore, \(j\) has a cartesian lifting
\(\mathbb j:\mathbb G\to \mathbb H\) by Lemma~\ref{lem:fib-lifted}. In particular,
\(\mathbb G\) is a functor.
\end{proof}
\end{thm}

In this work we only need internal monoidal categories that are strict.
In the same way as a strict monoidal category is a monoid in
\((\Cat,\mathbb 1, \times)\), an internal strict monoidal category is a
monoid in \((\Cat(\mathcal V),\mathbb I, \lotimes)\):

\begin{defi}[Internal monoidal category]\label{def:internal-monoidal-category}%
Let \((\mathcal V,1,\otimes)\) be a monoidal category with
finite limits in which the monoidal unit is the terminal object and let
\((\Cat(\mathcal V), \mathbb I, \lotimes)\) be the induced symmetric
monoidal category of internal categories in \(\mathcal V\). A strict
internal monoidal category \(\mathbb C\) is a monoid
\((\mathbb C, \varnothing, \odot)\) in
\((\Cat(\mathcal V), \mathbb I, \lotimes)\).
\end{defi}

\begin{rem}\label{rem:intmoncat-summary}
It may be useful to catalogue the different tensors. The first one is the cartesian product $\times$ of categories, with the help of which we define a monoidal product $\otimes$ on a particular category $\mathcal V$ and then lift it to a monoidal product $\lotimes$ on the category of categories internal in $\mathcal V$. This then allows us to define on an internal category $\mathbb C$ a tensor $\odot$, which we also call an \emph{internal tensor}:
\begin{gather*}
\otimes: \mathcal V \times \mathcal V\to \mathcal V \\
\lotimes: \Cat(\mathcal V)\times \Cat(\mathcal V)\to \Cat(\mathcal V)\\
\odot: \mathbb C\lotimes \mathbb C\to \mathbb C
\end{gather*}
\end{rem}

\begin{exa}
We see that the category \(\Fun\)
of Example~\ref{exle:nF-intmoncat} is an instance of Definition~\ref{def:internal-monoidal-category} if we choose \((\mathcal V,1,\otimes) = (\Nom,1,\ast)$ and $\mathbb C=\Fun$ and \(\odot=\uplus\).
\end{exa}

\begin{rem}
In the classical case where $\V=\Cat$ and both $\otimes$ and $\lotimes$ are the cartesian product, the interchange law for $\odot$ follows from $\odot$ being a functor. In the same way,  the interchange law for $\odot$ follows from $\odot$ being an \emph{internal} functor~\eqref{eq:def:internal-functor}
\[
\xymatrix@C=40pt{
(\mathbb C \lotimes \mathbb C)_2 \ar[r]^{\comp_{\mathbb C{\tiny \lotimes} \mathbb C}} \ar[d]_{\odot_2}
& (\mathbb C\lotimes\mathbb C)_1 \ar[d]^{\odot_1}
\\
\mathbb C_2\ar[r]_{\comp_\mathbb C}
& \mathbb C_1
}
\]
\end{rem}

\begin{exa}
In the category $(\Fun,\emptyset,\uplus)$ of finite sets of names and functions, see Example~\ref{exle:nF-intmoncat},  we have the interchange law
\[(\mathbb f\uplus \mathbb g)\then (\mathbb f'\uplus \mathbb g')
= (\mathbb f\then \mathbb f') \uplus (\mathbb g\then \mathbb g')\]
with the right-hand side being defined whenever the left-hand side is.
\end{exa}

% !TEX root =  calco2019-lmcs.tex

\section{Nominal PROPs and Nominal Monoidal Theories}

We introduce nominal $\PROP$s as particular instances of the internal categories of Section~\ref{sec:internal-monoidal}, taking care to explicate the definition in enough detail so that the reader can follow the rest of the paper without working through the details of the previous section. Nominal $\PROP$s in turn are presented by nominal monoidal theories ($\NMT$s) and it is with this notion that we start.  We also take the time to point out that sequential composition in NMTs is a binding operation and induces what we call diagrammatic $\alpha$-equivalence.

\hypertarget{nominal-monoidal-theories}{%
\subsection{Nominal monoidal
theories}\label{nominal-monoidal-theories}}

In analogy with symmetric monoidal theories (Section~\ref{sec:props-def}), a \emph{nominal monoidal theory} \(\label{NMT-def}(\Sigma,E)\) is given
by a set \(\Sigma\) of generators and a nominal set \(E\) of equations.
Each ordinary generator $\gamma : n \to m$ gives rise to a set of nominal generators:
The set $\nSigma$ of nominal generators consists of all
\([\boldsymbol a\rangle \gen \langle \boldsymbol b]: A \to B\)
where $A,B$ are sets of names of cardinality $n$ and $m$ respecitvely and \(\boldsymbol a, \boldsymbol b\) are lists,
enumerating the elements of $A,B$ (the enumerations list all elements and contain no repetitions).
The set of nominal generators
\(\nSigma\)
is closed under permutations:
\[\pi\cdot[\boldsymbol a\rangle \gen \langle \boldsymbol b]:\pi\cdot A\to\pi\cdot B = [\pi(\boldsymbol a)\rangle \gen \langle \pi(\boldsymbol b)] \tag{$\pi$\text{-def}}\label{cls-perm}.\]
The set of nominal terms or \(\)s is given by closing under the
operations of \autoref{fig:terms}, which should be compared with
\autoref{fig:smt-terms}.

\begin{figure}[h]
\begin{gather*}
\frac{\gen: m \to n \in \Sigma}{ [\boldsymbol a\rangle \gen \langle \boldsymbol b]: A \to B}
\qquad\qquad\quad
\frac{}{ id_a:\{a\}\to\{a\}}
\qquad\qquad\quad
\frac{}{\delta_{ab}:\{a\}\to\{b\}}
\\[2ex]
\frac{ t:A\to B\quad\quad t':A'\to B'}{ t\uplus t' :  A\uplus A'\to B\uplus B'}\qquad
\frac{ t:A\to B\quad\quad s:B\to C}{ t\hcomp s :  A \to C}
\qquad
\frac{ t:A\to B}{(a\ b)\,t :(a\ b)\cdot A\to (a\ b)\cdot B}
\end{gather*}
\caption{NMT Terms (\(\nTrm\))}\label{fig:terms}
\end{figure}

Every \(\NMT\) freely generates a monoidal category internal in nominal
sets by quotienting the generated terms by equations in \(E\), together
with the set  \(\label{NMTeq}\NMTeq\) of equations containing:

\begin{itemize}
\item
  the equations that state that \(\id\) and \(;\) obey the laws of a % chktex 40
  category
\item
  the equations stating that \(\id_\varnothing\) and \(\uplus\) are a
  monoid
\item
  the equations of an internal monoidal category of
  \autoref{fig:monoidal-category}
\item
  the equations of permutation actions of \autoref{fig:nominal-set}
\item
  the equations on the interaction of generators with bijections
  \(\delta\) of \autoref{fig:bijections}
\end{itemize}
Comparing $\SMTeq$ and $\NMTeq$, we find that the main difference between the equations in \autoref{fig:symmetric-monoidal-category} and in \autoref{fig:monoidal-category} is that the interchange law for \(\uplus\) is required to hold only if both sides are defined and that the two laws involving symmetries are replaced by the commutativity of \(\uplus\).

\begin{figure}[h]
\begin{align}
\tag{NMT-comm}%
\label{NMT-comm}
t\uplus s & = s \uplus t \qquad{}\\
\tag{NMT-ch}%
\label{NMT-ch}
(s \hcomp t) \uplus (u \hcomp v) &  = (s \uplus u) \hcomp (t \uplus v)
\end{align}
\caption{NMT Equations of \(\uplus\)}\label{fig:monoidal-category}
\end{figure}

\begin{figure}[h]
\begin{gather*}
(a\ b)id_x = id_{(a\ b)\cdot x}
\qquad\qquad
(a\ b)\delta_{xy} = \delta_{(a\ b)\cdot x \ (a\ b)\cdot y}
\qquad\qquad
(a\ b)\gen = (a\ b)\cdot \gen
\\[1ex]
(a\ b)(x\uplus y) = (a\ b)x\uplus (a\ b)y
\qquad\qquad
(a\ b)(x\hcomp y) = (a\ b)x\hcomp (a\ b)y
\end{gather*}
\caption{NMT Equations of the permutation actions}\label{fig:nominal-set}
\end{figure}

\begin{figure}[h]
\begin{gather*}
\delta_{aa} = \id_{a} \qquad\qquad
\delta_{ab}\hcomp\delta_{bc} = \delta_{ac}
\\[2ex]
\tag{NMT-left}%
\label{NMT-left}
\frac{[a_1,\hdots,a_i,\hdots,a_m\rangle \gen \langle \boldsymbol b]:\{a_i\}\uplus A\to B}{(\delta_{xa_i} \uplus \id_A)\,\then\, [a_1,\hdots,a_i,\hdots,a_m\rangle \gen \langle \boldsymbol b] = [a_1,\hdots,x,\hdots,a_m\rangle \gen \langle \boldsymbol b]}
\\[2ex]
\tag{NMT-right}%
\label{NMT-right}
\frac{[\boldsymbol a\rangle \gen \langle b_1,\hdots,b_i,\hdots,b_n]:A\to B\uplus\{b_i\}}{[\boldsymbol a\rangle \gen \langle b_1,\hdots,b_i,\hdots,b_n]\,\then\,(\id_{B} \uplus \delta_{b_ix})  = [\boldsymbol a\rangle \gen \langle b_1,\hdots,x,\hdots,b_n]}
\end{gather*}
\caption{NMT Equations of \(\delta\)}\label{fig:bijections}
\end{figure}

For terms to form a nominal set, we need equations between permutations
to hold, along with the equations of \autoref{fig:nominal-set} that specify
how permutations act on terms.

\begin{figure}[h]
\includegraphics[page=37, width=\linewidth]{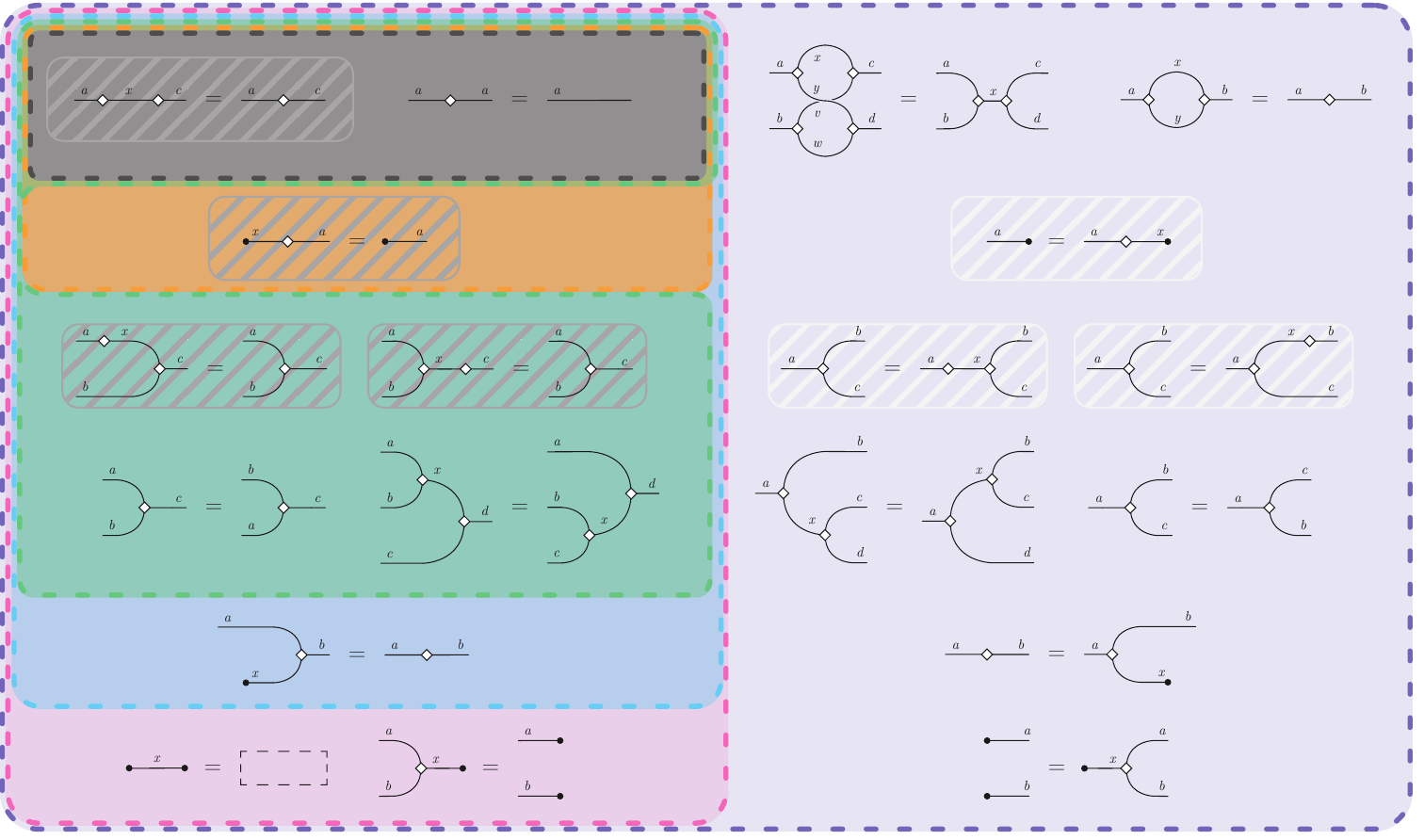}
\(\bijnbox\,\), \(\injnbox\,\), \(\surjnbox\,\), \(\funnbox\,\),
\(\pfunnbox\) and \(\relnbox\)
\caption{Nominal monoidal theories}\label{fig:nmt-theories}
\end{figure}

All the equations presented in the figures above are routine, with the
exception of the last two, specifying the interaction of renamings
\(\delta\) with the generators
\([\boldsymbol a\rangle \gen \langle \boldsymbol b]\in\Sigma\), which we
also depict in diagrammatic form:
\begin{center}
\begin{tabular}{ c c }
\includegraphics[page=33, width=5cm]{twists_new}
\quad\quad & \quad\quad
\includegraphics[page=34, width=5cm]{twists_new}\\
\end{tabular}
\end{center}
Instances of these rules can be seen in \autoref{fig:nmt-theories}, where
they are distinguished by a \stripbox{\strut striped} background.

\hypertarget{diagrammatic-alpha-equivalence}{%
\subsection{\texorpdfstring{Diagrammatic
\(\alpha\)-equivalence}{Diagrammatic \textbackslash alpha-equivalence}}\label{diagrammatic-alpha-equivalence}}

The equations of \autoref{fig:nominal-set} and \autoref{fig:bijections}
introduce a notion of \emph{diagrammatic \(\alpha\)-equivalence}, which
allows us to rename `internal' names and to contract renamings.

\begin{defi}\label{def:piA}
Two terms of a nominal monoidal theory are \(\alpha\)-equivalent if
their equality follows from the equations in \autoref{fig:nominal-set} and
\autoref{fig:bijections}.
\end{defi}

Every permutation \(\pi\) of names gives rise to bijective functions
\(\pi_A:A\to\pi[A]=\{\pi(a)\mid a\in A\}=\pi\cdot A\). Any such
\(\pi_A\), as well as the inverse \(\pi_A^{-1}\), are parallel
compositions of \(\delta_{ab}\) for suitable \(a,b\in\names\). In fact,
we have
\[\pi_A=\biguplus_{a\in A} \delta_{a\,\pi(a)}
\quad\quad \textrm{ and } \quad\quad
\pi^{-1}_A=\biguplus_{a\in A} \delta_{\pi(a)\,a}\]
We may therefore
use the \(\pi_A\) as abbreviations in terms. The following proposition is proved by induction on the structure of terms, see~\cite{balco:phd} for the details.

\begin{prop}
Let \(t:A\to B\) be a term of a nominal monoidal theory. The equations
in \autoref{fig:nominal-set} and \autoref{fig:bijections} entail that
\(\pi\cdot t = \pi_A^{-1}\hcomp t\hcomp \pi_B\). \[\xymatrix@C=50pt{
A\ar[r]^t \ar[d]_{\pi_A}& B\ar[d]^{\pi_B}\\
\pi[A]\ar[r]_{\pi\cdot t} & \pi[B]
}\]
\end{prop}

\begin{cor}
Let \(t:A\to B\) be a term of a nominal monoidal theory. Modulo the
equations of \autoref{fig:nominal-set} and \autoref{fig:bijections}, the
support of \(t\) is \(A\cup B\).

\begin{proof}
It follows from the proposition that \(\supp\,t\subseteq A\cup B\). For
the converse, suppose that there is \(x\in A\cup B\) and a support \(S\)
of \(t\) with \(x\notin S\subseteq A\cup B\). Choose a permutation
\(\pi\) that fixes \(S\) and maps \(x\) to some
\(\pi(x)\notin A\cup B\). Then either \(\pi\cdot A\not=A\) or
\(\pi\cdot B\not=B\), hence \(\pi\cdot t \not=t\), contradicting that
\(S\) is a support of \(t\).
\end{proof}
\end{cor}

The corollary shows that internal names are bound by sequential
composition. Indeed, in a composition
\(A\stackrel{t}{\to} C\stackrel{s}{\to}B\), the names in
\(C\setminus(A\cup B)\) do not appear in the support of \(t\hcomp s\).

%%%%%%%%%%%%%%%%%%%%
%%%%%%%%%%%%%%%%%%%%
\hypertarget{nominal-mathsfprops}{%
\subsection{Nominal PROPs}\label{nominal-mathsfprops}}

From the point of view of Section~\ref{sec:internal-monoidal}, a nominal
\(\mathsf{PROP}\), or \(\nPROP\) for short, is an internal strict
commutative monoidal category in \((\Nom,1,\ast)\) where
\begin{itemize}
\item objects are finite sets of names,
\item tensor is given by union of disjoint sets,
\item arrows contain all bijections,
\item permutations act on objects and bijections point-wise.
\end{itemize}
A functor between
\(\nPROP\)s is an internal strict monoidal functor that is the identity on objects and
bijections. For readers who skipped Section~\ref{sec:internal-monoidal}, we spell out the definition of $\nPROP$ explicitly:

\begin{defi}\label{def:nprop}
Given a countably infinite set $\names$ of `names', an \(\nPROP\) \((\mathbb C, \mathbb I, \uplus)\) consists of a set \(\mathbb C_0\) of `objects' and a set \(\mathbb C_1\) of `arrows' satisfying the following conditions.
\begin{itemize}
  \item
    \(\mathbb C_0\) is the set of finite subsets of
    \(\names\).
%    \item The support of a set of names \(A\) is the set itself and we define
%    \(\supp\,A = A\).
    \item $\mathbb C_0$ is equipped with a permutation action given by
    \(\pi \cdot A = \pi[A]=\{\pi(a) \mid a \in A\}\) for all finite permutations
    \(\pi:\names  \to  \names\).
    \item $\mathbb C$ is a category. We write
\(\then\) for its `sequential' composition (in the diagrammatic order).
  \item
    \(\mathbb C_1\) contains at least all bijections (`renamings')
    \(\pi_A:A\to\pi\cdot A\), see Definition~\ref{def:piA}, and is closed under the operation mapping an
    arrow \(f:A  \to  B\) to \(\pi \cdot f : \pi \cdot A  \to  \pi \cdot B\) defined as
    \(\pi \cdot f = (\pi_A)^{-1}\then f\then \pi_B\). %Such functions are referred to as finitely supported functions.
\item $\dom,\cod:\mathbb C_1\to\mathbb C_0$ preserve the permutation action.
\item $\uplus:\mathbb C_0\times\mathbb C_0 \to\mathbb C_0$ is the partial operation `union of disjoint sets'.
\item $\uplus:\mathbb C_1\times \mathbb C_1\to\mathbb C_1$ is a (partial) commutative and associative operation defined on  $\{(f , g) \in \mathbb C_1 \times \mathbb C_1 \mid \dom\, f \cap \dom\, g = \varnothing = \cod\, f \cap \cod\, g \}$ with the empty function as a unit.
\item $(f;f')\uplus (g;g')$ is defined whenever $(f\uplus g);(f'\uplus g')$ is and then both are equal.
\end{itemize}

\end{defi}

%From this definition on can deduce the following.

\begin{rem}\label{rem:nprop}
\
\begin{itemize}
\item
  A nominal \(\PROP\) has a nominal set of objects and a nominal set of
  arrows.

  \item

  The support $\supp(A)$ of an object \(A\) is \(A\) and the support of an arrow
  \(f:A\to B\) is \(A\cup B\). In particular,
  \(\supp\,(f\then g) = \dom\, f\cup \cod\, g\). In other words, nominal
  \(\PROP\)s have diagrammatic \(\alpha\) equivalence.
\item
  There is a category \(\nPROP\) that consists of nominal \(\PROP\)s
  together with functors that are the identity on objects and strict
  monoidal and equivariant.
\item
  Every $\NMT$ presents a \(\nPROP\). Conversely, every \(\nPROP\) is
  presented by at least one $\NMT$ given by all terms as generators and all
  equations as relations.
\end{itemize}
\end{rem}

\hypertarget{sec:examples}{%
\subsection{Examples}\label{sec:examples}}

We present as
examples those \(\NMT\)s that correspond to the \(\SMT\)s of
\autoref{fig:smt-theories}. The significant differences between
\autoref{fig:smt-theories} and \autoref{fig:nmt-theories} are that wires now
carry labels and that there is a new generator
\raisebox{-1pt}{\includegraphics[page=43, width=20mm]{twists_new}}
which allows us to change the label of a wire. Moreover, in the nominal
setting, rules for wire crossings are not needed.

%The first example lists presentations of nominal monoidal theories for
%the nominal monoidal categories of finite sets and functions,
%injections, surjections, partial functions and relations, respectively.
%
\begin{exa}%
\label{ex:nmt-presentations}
We spell out the details of \autoref{fig:nmt-theories}. The category of finite nominals sets and

\begin{itemize}
\item
  bijections is presented by the empty signature and equations.
\item
  injections is presented by
  \(\Sigma_i=\{\eta_a:\varnothing \to \{a\} \mid a \in \mathcal N\,\}\)
  and \(E_i=\emptyset\). The equations

  \begin{center}
  \includegraphics[page=5, width=5cm]{twists_new}
  \end{center}

  \vspace{-3ex}
\ \\
follow from those of \autoref{fig:bijections}.
\item
  surjections is presented by
  \(\Sigma_s=\{\mu_{abc}:\{a,b\} \to \{c\} \mid a,b,c \in \mathcal N\,\}\)
  and equations \(E_s\) are
  \((\mu_{abx} \uplus id_c) \circ \mu_{cdx} = (\mu_{bcx} \uplus id_a) \circ \mu_{adx}\),
  presented graphically as
  \[
  \includegraphics[page=12, width=5cm]{twists_new}
  \]
\item
  functions has \(\Sigma_f=\Sigma_i \cup \Sigma_s\) and equations
  \(E_f\) are \(E_i\cup E_s\) plus
  \((id_a \uplus \eta_x) \circ \mu_{abx} = \delta_{ab}\)
\item
  partial functions has
  \(\Sigma_{pf}=\Sigma_f \cup \{\hat\eta_a:\{a\} \to \varnothing \mid a \in \mathcal N\,\}\)
  and equations \(E_{pf}\) are \(E_f\) plus
  \(\eta_x \circ \hat\eta_x = \varepsilon\) and
  \(\mu_{abx} \circ \hat\eta_x = \hat\eta_a \uplus \hat\eta_b\), shown
  below

  \begin{center}
  \raisebox{0.5cm}{\includegraphics[page=6, width=5cm]{twists_new}}
  \qquad \includegraphics[page=16, width=5cm]{twists_new}
  \end{center}
\item
  relations has
  \(\Sigma_{r}=\Sigma_{pf} \cup \{\hat\mu_{abc}:\{a\} \to \{b,c\} \mid a,b,c \in \mathcal N\,\}\),
  and equations \(E_r\) are \(E_{pf}\) plus the following

  \begin{center}
  \includegraphics[page=13, width=5cm]{twists_new}
  \qquad\qquad
  \raisebox{0.5cm}{\includegraphics[page=19, width=5cm]{twists_new}}
  \includegraphics[page=9, width=5cm]{twists_new}\qquad
  \includegraphics[page=18, width=5cm]{twists_new}
  \includegraphics[page=11, width=5cm]{twists_new}
  \end{center}
\end{itemize}
\end{exa}

% !TEX root =  calco2019-lmcs.tex

\section{Equivalence of Nominal and Symmetric Monoidal Theories}\label{sec:eq-theories}

We should be able to switch easily between a notion of ordered names on the one hand and a notion of unordered abstract names  on the other. This
intuition is reinforced by putting Figures~\ref{fig:smt-theories} and~\ref{fig:nmt-theories} next to each other. A careful investigation suggests that there is a general procedure to automatically translate one into the other. This section will give such translations and prove that these translations are inverse to each other and preserve completeness. This yields a tool to derive completeness of an $\NMT$ from the completeness of the corresponding $\SMT$ and vice versa.

In Section~\ref{sec:theories} we define operations $\Prop : \SMT \to \PROP$ and $\nProp : \NMT \to \nPROP$ quotienting theories to their represented categories. Sections~\ref{sec:nom-translation}-\ref{sec:equiv} are devoted to showing that the categories $\PROP$ and $\nPROP$ are equivalent. Section~\ref{sec:nfNmt} defines the translation $\nfNmt:\SMT\to\NMT$ and Section~\ref{sec:NMTcompleteness} shows that $\nfNmt(\mathcal S)$ is complete if $\mathcal S$ is complete. Sections~\ref{sec:nfSmt}-\ref{sec:SMTcompleteness} establish the analogous result in the other direction.

\subsection{Interpreting (nominal) monoidal theories as (nominal) PROPs}\label{sec:theories}

We start by formalising the operation that maps a theory to the category it represents.
Given a theory \(\langle \Sigma , E\rangle \) of string diagrams,  where \(\Sigma \) is the set of generators and
\(E \subseteq \Trm(\Sigma ) \times \Trm(\Sigma )\) is the set of equations, the  operation \(\Prop : \SMT \to \PROP\) takes the signature \(\langle \Sigma , E\rangle \)
to the category of \(\SMT\) terms, quotiented by the equations of \(E\).

\begin{defi}\label{def:Prop-def}
The functor
\(\Prop : \SMT \to \PROP\) is defined as
\[\Prop\, \langle \Sigma , E\rangle = \Trm(\Sigma )/\Cl(E\cup \SMTeq)\]
\end{defi}

\begin{figure}[h]
\begin{center}
\begin{gather*}
\frac{}{ s = s \in \Cl(E)}
\qquad\qquad\quad
\frac{s = t \in \Cl(E)}{ t = s \in \Cl(E)}
\\[2ex]
\frac{s = t \in \Cl(E) \quad\quad t = u \in \Cl(E)}{ s= u \in \Cl(E)}
\qquad\qquad\quad
\frac{s = s' \in \Cl(E) \quad\quad t = t' \in \Cl(E)}{ s * t = s' * t' \in \Cl(E)}\qquad
\end{gather*}
\end{center}
\caption{Closure operator}\label{fig:cl-op}
\end{figure}

This definition uses the closure operator \(\Cl\), defined in
\autoref{fig:cl-op}, which is the usual deductive closure of equational
logic. We have \(* \in \{\hcomp, \,\oplus \,\}\) for equations
on \(\Trm\)s and for equations on \(\nTrm\)s we have
\(* \in \{\hcomp, \,\uplus \,\}\) along with an additional rule for
permutations: \[\frac{s = t \in \Cl(E)}{ \pi \cdot s = \pi \cdot t \in \Cl(E)}\]

We have a similar construction for \(\NMT\)s, where we define a functor
\(\nProp : \NMT \to \nPROP\):

\begin{defi}\label{def:nProp-def}
The functor \(\nProp : \NMT \to \nPROP\) is defined as
\[\nProp\,\langle \Sigma ,E\rangle = \nTrm(\Sigma )/\Cl(E \cup \NMTeq)\]
\end{defi}

We will need later that homomorphisms of term-algebras, see Sections~\ref{sec:NMTcompleteness} and~\ref{sec:SMTcompleteness}, preserve equations.

\begin{lem}\label{lem:nfNmtCl}
Given a set of equations \(X \subseteq \nTrm(A) \times \nTrm(A)\), or
\(X \subseteq \Trm(A) \times \Trm(A)\), and a homomorphism
\(f : \nTrm(A) \to \nTrm(B)\), or \(f : \Trm(A) \to \Trm(B)\), we have
\[f[\Cl(X)] \subseteq \Cl(f[X]).\]
%\begin{proof}
%The statement above is equivalent to
%\(\forall (s,t) \in \Cl(X)\,.\ \,(f(s),f(t)) \in \Cl(f[X])\).\\
%Then, by induction on the formation rules of the set \(\Cl(X)\), we have
%the following cases:
%
%\begin{itemize}
%\item
% If \((s,t) \in X\), then \((f(s),f(t)) \in f[X]\), by definition and
% therefore, \((f(s),f(t)) \in \Cl(f[X])\).
%\item
% If \((s,t) \in \Cl(X)\), by reflexivity, symmetry or transitivity, then
% by IH \((f(s),f(t)) \in \Cl(f[X])\).
%\item
% If \((s,t) \in \Cl(X)\), by congruence of \(\hcomp\) or \(\uplus \) or
% permutation, the result follows by IH and the fact that \(f\) is a
% homomorphism, e.g.:
%
% For \((s \uplus t,s' \uplus t') \in \Cl(X)\), by IH, we have
% \[(f(s),f(s')) \in \Cl(f[X]) \text{ and } (f(t),f(t')) \in \Cl(f[X])\]
% then we also have \[(f(s) \uplus f(t),f(s') \uplus f(t')) \in \Cl(f[X])\] and
% since we know \(f\) is a homomorphism, we have
% \(f(s \uplus t) = f(s) \uplus f(t)\), thus \[(f(s \uplus t),f(s' \uplus t')) \in \Cl(f[X])\]
%\end{itemize}
%\end{proof}
\end{lem}

\hypertarget{sec:nom-translation}{%
\subsection{Embedding PROPs into
nominal PROPs}\label{sec:nom-translation}}

In this section, we start proving the equivalence of the categories $\PROP$ and $\nPROP$ by embedding ordinary $\PROP$s into $\nPROP$s. This is achieved in the following manner. Given
an ordinary diagram \(f : n \to m\),

\begin{center}
\includegraphics[page=52, width=7cm]{twists_new_thesis}
\end{center}

we create ``boxed'' nominal versions \([\boldsymbol a\rangle f\langle \boldsymbol b]\), where
\(\boldsymbol a = [a_1,\hdots,a_n]\) and
\(\boldsymbol b = [b_1,\hdots,b_m]\) are lists of pairwise distinct names:
\begin{center}
\includegraphics[page=53, width=7cm]{twists_new_thesis}
\end{center}

For ``boxing''  to preserve the relevant structure, we have to ensure, in particular, that the symmetric monoidal tensor of a $\PROP$s is mapped to the commutative tensor of $\nPROP$s, and that sequential composition and identities are preserved:

\begin{center}
\includegraphics[page=54, width=9cm]{twists_new_thesis}

\includegraphics[page=55, width=7cm]{twists_new_thesis}

\includegraphics[page=56, width=7cm]{twists_new_thesis}
\end{center}

We give the details of this construction in the proposition below, using the following notational conventions.

\begin{notation}%
\label{boxdia}
\leavevmode
\begin{itemize}
\item If $\boldsymbol a = [a_1,\ldots a_n]$ is a list, then $\underline {\boldsymbol a}=set(\boldsymbol a)=\{a_1,\ldots a_n\}$.
\item The elements of $[a_1,\ldots a_n]$ will always be pairwise distinct and we say that $\boldsymbol a$ is an \emph{enumeration} of $set(a)$.
\item Given enumerations $\boldsymbol a$ and $\boldsymbol a'$ with $\underline{\boldsymbol a}=\underline{\boldsymbol a'}$ we abbreviate the bijection mapping $i\mapsto j$ whenever $a_i=a'_j$  as
\[\langle \boldsymbol a | \boldsymbol a'\rangle\]
\item Given enumerations $\boldsymbol a$ and $\boldsymbol b$ of the same length, we write
\[[ \boldsymbol a | \boldsymbol b]\]
for the bijection  $\biguplus \delta_{a_ib_i}$ mapping $a_i\mapsto b_i$.
\item Given two enumerations $\boldsymbol a$ and $\boldsymbol b$, we write $\boldsymbol a \concat \boldsymbol b$ for their concatenation (as lists).
\end{itemize}
\end{notation}

\begin{prop}%
\label{prop:NOM}
For any $\PROP$ $\mathcal S$, there is an $\nPROP$
\[\NOM(\mathcal S)\]
that has an arrow $[\boldsymbol a\rangle f\langle\boldsymbol b]:\underline{\boldsymbol a}\to \underline{\boldsymbol b}$ for all arrows $f:\underline n\to \underline m$ of $\mathcal S$ and for all lists $\boldsymbol a = [a_1,\ldots a_n]$ and $\boldsymbol b = [b_1,\ldots b_m]$ of pairwise distinct names. These arrows are subject to the set \[\NOMeq\] consisting of  the equations
\begin{align*}
[\boldsymbol a\rangle f\hcomp g\langle\boldsymbol c]
&=
[\boldsymbol a\rangle f\langle\boldsymbol b]\hcomp[\boldsymbol b\rangle g\langle\boldsymbol c]
&\tag{NOM-1}\label{NOM-1}
\\
[\boldsymbol a \concat \boldsymbol c\rangle f \oplus g\langle\boldsymbol b \concat \boldsymbol d]
&=
[\boldsymbol a\rangle f \langle\boldsymbol b]\uplus[\boldsymbol c\rangle g\langle\boldsymbol d]
&\tag{NOM-2}\label{NOM-2}
\\
[\boldsymbol a\rangle \id \langle \boldsymbol b]
& = [\boldsymbol a| \boldsymbol b]
&\tag{NOM-3}\label{NOM-3}
\\
[\boldsymbol a\rangle\, \langle\boldsymbol b | \boldsymbol b'\rangle \hcomp f\, \langle\boldsymbol c]\,
& =
[\boldsymbol a | \boldsymbol b]\hcomp [ \boldsymbol b'\rangle f \langle\boldsymbol c]
&\tag{NOM-4}\label{NOM-4}
\\
[\boldsymbol a\rangle\, f \hcomp \langle\boldsymbol b | \boldsymbol b'\rangle\, \langle\boldsymbol c]\,
& =
[ \boldsymbol a\rangle f \langle\boldsymbol b] \hcomp [\boldsymbol b' | \boldsymbol c]
&\tag{NOM-5}\label{NOM-5}
\end{align*}

Then, $\NOM:\PROP\to\nPROP$ is a functor mapping an arrow of $\PROP$s $F:\mathcal S\to\mathcal S'$ to an arrow of $\nPROP$s
$\NOM(F):\NOM(\mathcal S)\to\NOM(\mathcal S')$ defined by
$\NOM(F)([\boldsymbol a \rangle\, g\,\langle\boldsymbol b]) = [\boldsymbol a \rangle\, Fg\,\langle\boldsymbol b].\label{NOM-F}$
\end{prop}

In the proposition above, $[\boldsymbol a\rangle f\langle\boldsymbol b]$ is a purely formal generator. Of course, the intuition is that $[\boldsymbol a\rangle f\langle\boldsymbol b]$ represents a composition of maps $[\boldsymbol a\rangle \then f\then \langle\boldsymbol b]$. This can be made explicit as in the following example.

\begin{exa}%
\label{ex:nFNOMiso}
$\nF$ is isomorphic to $\NOM(\FunSkel)$ with the isomorphism $G : \NOM(\FunSkel) \to  \nF$ being defined as
\[G ([\boldsymbol a \rangle  f \langle \boldsymbol b ]) = \semdia{\boldsymbol a }{f}{\boldsymbol b }\]

where the `semantic box' $\semdia{-}{-}{-}$ translates an arrow $f:n\to m$  by pre-composing with $\overrightarrow{\boldsymbol a } : A \to  n$ and post-composing with $\overrightarrow{\boldsymbol b }^{-1} : m \to  B$, where $\overrightarrow{\boldsymbol a }$ is the bijection between $\underline{\boldsymbol a} $ and $n$ given by the ordering of the list $\boldsymbol a $. In other words, we have $\semdia{\boldsymbol a }{f}{\boldsymbol b } \ \eqdef \ \overrightarrow{\boldsymbol a } \hcomp f \hcomp \overrightarrow{\boldsymbol b }^{-1}$.
\end{exa}

\hypertarget{sec:ord-translation}{%
\subsection{Embedding nominal PROPs into PROPs}\label{sec:ord-translation}}

We can embed $\nPROP$s into $\PROP$s in much the same fashion. We write this translation in one-dimensional syntax as \(\langle \boldsymbol a]\,f\,[\boldsymbol b\rangle \) and in two-dimensional syntax as

\begin{center}
\includegraphics[page=58, width=7cm]{twists_new_thesis}
\end{center}

For the next proposition we use again Notation~\ref{boxdia}.

\begin{prop}%
\label{prop:ORD}
For any $\nPROP$ $\mathcal T$ there is a $\PROP$
\[\ORD(\mathcal T)\]
that has
an arrow $\langle\boldsymbol a]f[\boldsymbol b\rangle\in\ORD(\mathcal T)$
for each arrow $f:A\to B$ of $\mathcal T$ and for all lists $\boldsymbol a = [a_1,\ldots a_n]$ and $\boldsymbol b = [b_1,\ldots b_m]$ of pairwise distinct names. These arrows are subject to the set  \[\ORDeq\] of equations
\begin{align*}
\langle\boldsymbol a]\,f\hcomp g\,[\boldsymbol c\rangle
&=
\langle\boldsymbol a]\, f \,[\boldsymbol b\rangle\hcomp\langle\boldsymbol b]\, g\,[\boldsymbol c\rangle
&\tag{ORD-1}\label{ORD-1}
\\
\langle\boldsymbol a_f \concat \boldsymbol a_g]\,f\uplus g\,[\boldsymbol b_f \concat \boldsymbol b_g\rangle
&=
\langle\boldsymbol a_f]\,f\,[\boldsymbol b_f\rangle \oplus \langle\boldsymbol a_g]\,g\,[\boldsymbol b_g\rangle
&\tag{ORD-2}\label{ORD-2}
\\
\langle\boldsymbol a]\,\id\,[\boldsymbol a\rangle
& = \id
&\tag{ORD-3}\label{ORD-3}
\\
\langle\boldsymbol a] \, [\boldsymbol a' | \boldsymbol b]\hcomp f\,  [\boldsymbol c\rangle
& =
\langle\boldsymbol a | \boldsymbol a'\rangle \hcomp \langle \boldsymbol b]\, f \,[\boldsymbol c\rangle
&\tag{ORD-4}\label{ORD-4}
\\
\langle\boldsymbol a]\, f \hcomp [\boldsymbol b | \boldsymbol c] \, [\boldsymbol c'\rangle
& =
\langle \boldsymbol a]\, f \,[\boldsymbol b\rangle \hcomp \langle\boldsymbol c | \boldsymbol c'\rangle
&\tag{ORD-5}\label{ORD-5}
\\
\langle\boldsymbol a]\, f \, [\boldsymbol b\rangle
& =
\langle \pi\cdot \boldsymbol a]\, \pi\cdot f \,[\pi\cdot \boldsymbol b\rangle
&\tag{ORD-6}\label{ORD-6}
\end{align*}

Then, $\ORD$ is a functor mapping an arrow of $\nPROP$s $F:\mathcal T\to\mathcal T'$ to an arrow of $\PROP$s
$\ORD(F):\ORD(\mathcal T)\to\ORD(\mathcal T')$ defined by

\[\ORD(F)(\langle\boldsymbol a ]\, f \,[\boldsymbol b\rangle) = \langle\boldsymbol a ]\,  Ff  \,[\boldsymbol b\rangle\label{ORD-F}.\]
\end{prop}

In the proposition above, similarly to Proposition~\ref{prop:NOM}, the notation $\langle\boldsymbol a]f[\boldsymbol b\rangle$ denotes formal generators. The intended meaning as a composition of three functions is made clear by the next example for which we borrow notation from Example~\ref{ex:nFNOMiso}.

\begin{exa}%
\label{ex:FunSkelORDiso}
The category $\FunSkel$  is isomorphic to $\ORD(\nF)$, with the isomorphism $G : \ORD(\nF) \to  \FunSkel$ given, for all  $f : A \to  B$, as
\[G (\langle \boldsymbol a ] f [\boldsymbol b \rangle ) = \sembox{\boldsymbol a }{f}{\boldsymbol b }\]

where the `semantic diamond' $\sembox{-}{-}{-}$ translates an arrow $f \in   \nF$ into an arrow in $\FunSkel$, defined as $\sembox{\boldsymbol a  }{f}{\boldsymbol b  } \ \eqdef\  \overrightarrow{\boldsymbol a  }^{-1} \hcomp f \hcomp \overrightarrow{\boldsymbol b  }$. The detailed verifications are straightforward and can be found in~\cite[Example 7.26]{balco:phd}.
\end{exa}

\hypertarget{sec:equiv}{%
\subsection{Equivalence of PROPs and nominal PROPs}\label{sec:equiv}}

We show that $\PROP$ and $\nPROP$ are equivalent. More precisely, the functors $\ORD$ from Proposition~\ref{prop:ORD} and $\NOM$ from Proposition~\ref{prop:NOM} form an equivalence of categories.

\begin{prop}
\normalfont For each $\PROP$ $\mathcal S$, there is an isomorphism of $\PROP$s, natural in $\mathcal S$,
\[\Delta :
\mathcal S \to \ORD(\NOM(\mathcal S))\]
mapping $f \in \mathcal S$ to $\langle\boldsymbol a][\boldsymbol a\rangle\, f\,\langle\boldsymbol b][\boldsymbol b\rangle$ for some choice of $\boldsymbol a, \boldsymbol b$.

\begin{proof}For a proof see~\cite[Proposition 7.33]{balco:phd}.\end{proof}
\end{prop}

\begin{prop}
\normalfont For each $\nPROP$ $\mathcal T\,$, there is an isomorphism of $\nPROP$s, natural in $\mathcal T$,
\[\NOM(\ORD(\mathcal T))\to\mathcal T\]
mapping the $[\boldsymbol c\rangle\langle \boldsymbol a]\,f\,[\boldsymbol b\rangle\langle\boldsymbol d]$ generated by an $f:\underline{\boldsymbol a}\to\underline{\boldsymbol b}$ in $\mathcal T$ to $[\boldsymbol c|\boldsymbol a] \hcomp f \hcomp [\boldsymbol b|\boldsymbol d]$.

% \begin{proof}
% We define a converse ${\Delta_{\mathsf n} : \mathcal T\to \NOM(\ORD(\mathcal T))}$ mapping ${f:\underline{\boldsymbol a}\to\underline{\boldsymbol b}}$ to $[\boldsymbol a\rangle\langle \boldsymbol a] \,f\,[\boldsymbol b\rangle\langle\boldsymbol b]$ for some choice of $\boldsymbol a, \boldsymbol b$.

% The verification that $\Gamma_{\mathsf n} (\Delta_{\mathsf n} (f)) = f$ for any $f$ can be found in~\cite{balco:phd}.
% \end{proof}

\begin{proof}For a proof see~\cite[Proposition 7.34]{balco:phd}.\end{proof}
\end{prop}

Since the last two propositions provide an isomorphic unit and counit of an adjunction, we obtain

\begin{thm}%
\label{thm:equivalence}
\normalfont The categories $\PROP$ and $\nPROP$ are equivalent.
\end{thm}

\begin{rem}
If we generalise the notion of $\PROP$ from MacLane~\cite{maclane} to Lack~\cite{lack}, in other words, if we drop the last equation of \autoref{fig:symmetric-monoidal-category} expressing the naturality of symmetries, we still obtain an adjunction, in which $\NOM$ is left-adjoint to $\ORD$. Nominal $\PROP$s then are a full reflective subcategory of ordinary $\PROP$s. In other words, the (generalised) $\PROP$s $\mathcal S$ that satisfy naturality of symmetries are exactly those for which $\mathcal S\cong \ORD(\NOM(\mathcal S))$.
Example~\ref{ex:nfNmt} shows in more detail how naturality of symmetries is a consequence of the commutativity of the nominal tensor.
\end{rem}

\hypertarget{sec:nfNmt}{%
\subsection{Translating SMTs into NMTs}\label{sec:nfNmt}}

We give a formal definition of the translation of the ordinary theories of Figure~\ref{fig:smt-theories} to the nominal theories of Figure~\ref{fig:nmt-theories}.

First let us recall that the composition of $\Prop$ from Remark~\ref{def:Prop-def} with the functor $\NOM$ from Proposition~\ref{prop:NOM}, gives us the following interpretation of a $\PROP$ as an $\nPROP$:
Given an \(\SMT\) \(\langle \Sigma ,E\rangle \), we can generate an \(\nPROP\), by simply
taking all the \(\SMT\)-terms over \(\Sigma \), as generators (taking
\(\Trm(\Sigma )\) to \(\nTrm(\Trm(\Sigma ))\)) and taking \(\boxTrm(E) \cup \NOMeq\) as equations, where:

\begin{itemize}
\item
 \(\label{box-def}\boxTrm(E) = \{ [\boldsymbol a\rangle f\langle \boldsymbol b] = [\boldsymbol a\rangle g\langle \boldsymbol b] \mid f = g \in E\}\)
\item
 \(\NOMeq\) are the equations from Proposition~\ref{prop:NOM}
\end{itemize}

\noindent
Whilst this construction gives us a way of embedding equivalence classes of ordinary string diagrams into equivalence classes of nominal ones, it does not  answer the question of how to translate the axioms defining an $\SMT$ into the axioms of the corresponding $\NMT$.

If we recall the definition of an $\NMT$, we see that the signature of a nominal theory consists of a set of ordinary generators $\Sigma$ and set of equations over $\nTrm(\Sigma)$. Thus, given the ordinary signature of an $\SMT$, with generators $\Sigma$ and the set of equations $E \subseteq\Trm(\Sigma)\times \Trm(\Sigma)$,  we need to obtain an $E'\subseteq \nTrm(\Sigma)\times\nTrm(\Sigma)$ such that any equivalence class induced by $E'$ and the equations of $\NOM$ (due to the ordinary diagrams being embedded in nominal ones) are mirrored by $E$.

Intuitively, we translate equations of $E$, by first embedding them inside a `nominal box' as a whole and then use the rules of $\NOM$ to recursively normalise all “sub-diagrams” into nominal ones (see Example~\ref{ex:nfNmt}). When we hit the ``base case'', i.e.\ a ``boxed'' generator from $\Sigma$, we simply replace it with a corresponding nominal generator:

\begin{center}
\includegraphics[page=65, width=7cm]{twists_new_thesis}
\end{center}

Below, we use the notation \(\underline{\gamma }\) to
highlight the difference between an element \(\gamma \) of \(\Sigma \) and the string diagram \(\underline{\gamma } \in \Trm(\Sigma )\) as in the blue box above.

The definition \(\nfNmt : \nTrm(\Trm(\Sigma )) \to \nTrm(\Sigma )\) is thus straightforward:
%
%\medskip
%Whilst the construction above interpretes
%\(\SMT\)s as \(\nPROP\)s, we want to have a direct way of translating \(\SMT\)s
%into \(\NMT\)s. We do this by defining the function
%$$\nfNmt : \nTrm(\Trm(\Sigma )) \to \nTrm(\Sigma )$$
%via
\begin{align*}& \nfNmt\, ([\boldsymbol a \rangle \underline{\gamma } \langle \boldsymbol b ]) = [\boldsymbol a \rangle \gamma \langle \boldsymbol b ] \text{ where } \gamma \in \Sigma \\
 & \nfNmt\, ([a\rangle id \langle b]) = \delta_{ab}\\
 & \nfNmt\, ([ab\rangle \sigma \langle cd]) = [ab|dc]\\
 & \nfNmt\, ([\boldsymbol a \rangle f \hcomp g \langle \boldsymbol c ]) = \nfNmt\, ([\boldsymbol a \rangle f \langle \boldsymbol b ]) \hcomp \nfNmt\, ([\boldsymbol b \rangle g \langle \boldsymbol c ])\\
 & \nfNmt\, ([\boldsymbol a \concat \boldsymbol b \rangle f \oplus g \langle \boldsymbol c \concat \boldsymbol d ]) = \nfNmt\, ([\boldsymbol a \rangle f \langle \boldsymbol c ]) \uplus \nfNmt\, ([\boldsymbol b \rangle g \langle \boldsymbol d ]) \\
 & \nfNmt\, (id_{a}) = id_{a}\\
 & \nfNmt\, (\delta_{ab}) = \delta_{ab}\\
 & \nfNmt\, (f \hcomp g) = \nfNmt\, (f) \hcomp \nfNmt\, (g)\\
 & \nfNmt\, (f \uplus g) = \nfNmt\, (f) \uplus \nfNmt\, (g)\\
 & \nfNmt\, (\pi \cdot f) = \pi \cdot \nfNmt\, (f)
\end{align*}

%In the graphical notation the first equation appears as
%\begin{center}
%\includegraphics[page=65, width=7cm]{twists_new_thesis}
%\end{center}

%This picture can be read as mapping the ordinary generator \(\mu \) to the nominal generator \([a,b\rangle \mu \langle c]\).

\begin{defi}\label{Nmt-def}
We define \(\Nmt : \SMT \to \NMT\) as
\[\Nmt\,\langle \Sigma ,E\rangle = \langle \Sigma ,\nfNmt (\boxTrm(E))\rangle \] where we extend the function
\(\nfNmt\) on a set of equations in the obvious way:
\[\nfNmt\,(E) = \{ \nfNmt(f) = \nfNmt(g) \mid f = g \in E\}\]
\end{defi}

We now return to \autoref{fig:smt-theories} and \autoref{fig:nmt-theories} and
show at the hand of an example how, by applying \(\Nmt\) to the equations in
\autoref{fig:smt-theories}, we obtain the equations in \autoref{fig:nmt-theories}.

\begin{exa}\label{ex:nfNmt}
In this example, we illustrate the translation of a rule of an \(\SMT\)
into the corresponding rule of an \(\NMT\) via \(\nfNmt\). The diagram
below shows the application of \(\nfNmt\) to both sides of an equation
in the \(\SMT\) theory of surjections.

\begin{center}
\includegraphics[page=63, width=10cm]{twists_new_thesis}
\end{center}

The next diagram illustrates that the equation~\eqref{SMT-nat}
is subsumed by the equations of \(\NMT\), namely by~\eqref{NMT-comm}.

\begin{center}
\includegraphics[page=64, width=12cm]{twists_new_thesis}
\end{center}
\end{exa}

%%%%%%%%%%%%%%%%%%%%%%%%%%%%%%%%%%%%%%%%
%%%%%%%%%%%%%%%%%%%%%%%%%%%%%%%%%%%%%%%%
\hypertarget{sec:NMTcompleteness}{%
\subsection{Completeness of NMTs}\label{sec:NMTcompleteness}}

We show how to prove the completeness of $\Nmt(\Sigma,E)$ from the completeness of $\langle \Sigma,E\rangle$.
The central observation is that the diagram in \autoref{fig:nmt-compl-sq} commutes up to isomorphism.

\floatstyle{plain}
\restylefloat{figure}

\begin{figure}
\[\xymatrix{
% Level 1
\SMT
\ar[0,4]|-{\ \Prop\ }
\ar[5,0]_{\NmtSubscript}
& & & &
\PROP
\ar[5,0]^{\NOMSubscript}
\\
% Level 2
& \langle \Sigma , E\rangle
\ar@{ |->}[0,2]
\ar@{ /->}`d[dddr][dddrr]
& & \Prop\, \langle \Sigma , E\rangle
\ar@{ /->}[dd]
 &
\\
% Level 3
 & & & &
\\
% Level 4
 & & &
\NOM(\Prop\, \langle \Sigma , E\rangle )
\ar@/^1.2pc/[d]^{\nfNmtSubscript}
\\
% Level 5
 & & &
\nProp(\Nmt\, \langle \Sigma , E\rangle )
\ar@/^1.2pc/[u]^{\injNmtSubscript}
\\
% Level 6
\NMT
\ar[0,4]|-{\ \nProp\ }
& & & &
\nPROP
%\ar@<1ex>[0,1]^-{\dom}
%\ar@<-1ex>[0,1]_-{\cod}
%&
%\ \Fun_0
}\]
\caption{Completing the square}\label{fig:nmt-compl-sq}
\end{figure}

%\floatstyle{boxed}
%\restylefloat{figure}

First, we define the homomorphism of term-algebras
\(\label{injNmt-def} \injNmt : \nTrm(\Sigma ) \to \nTrm(\Trm(\Sigma ))\):
\begin{align*}& \injNmt([\boldsymbol a \rangle \gamma \langle \boldsymbol b ]) = [\boldsymbol a \rangle \underline{\gamma }\langle \boldsymbol b ] \text{ where } \gamma \in \Sigma \\
 & \injNmt(id_{a}) = id_{a}\\
 & \injNmt(\delta_{ab}) = \delta_{ab}\\
 & \injNmt(f \hcomp g) = \injNmt(f) \hcomp \injNmt(g)\\
 & \injNmt(f \uplus g) = \injNmt(f) \uplus \injNmt(g)\\
 & \injNmt(\pi \cdot f) = \pi \cdot \injNmt(f)\\
\end{align*}
The only interesting case is the one of  a
nominal generator $[\boldsymbol a \rangle \gamma \langle \boldsymbol b ]$, which gets turned into an ordinary string diagram $\underline\gamma$,
embedded in a nominal diagram $[\boldsymbol a \rangle \underline{\gamma }\langle \boldsymbol b ]$.

\begin{lem}\label{lem:completeSquareNMT}
The diagram in \autoref{fig:nmt-compl-sq} commutes up to isomorphism.

\begin{proof}
We want to show that the two maps \(\nfNmt\) and \(\injNmt\) are
isomorphisms. By definition, both \(\nfNmt\) and \(\injNmt\) are
homomorphisms between the term algebras. To verify that they are well-defined, that is, that they respect
equivalence classes of \(\nTrm\)s, we need to check that they preserve
the equations:

\begin{itemize}
\item
 For the map \(\injNmt\) we have to show
 \[\injNmt[\Cl(\nfNmt [\boxTrm[E]] \cup \NMTeq)] \subseteq \Cl(\Cl(\boxTrm[E \cup \SMTeq]) \cup \NOMeq \cup \NMTeq)\]
 In fact, by Lemma~\ref{lem:nfNmtCl}, it suffices to check that
 \(\injNmt[\nfNmt [\boxTrm[E]]] \subseteq \Cl(\boxTrm[E] \cup \NMTeq)\) and
 \(\injNmt[\NMTeq] \subseteq \Cl(\boxTrm[E] \cup \NMTeq)\). The first inequality
 follows immediately from the fact that
 \(\injNmt[\nfNmt [\boxTrm[E]] = \boxTrm[E]\). The second inequality
 follows straightforwardly.
 %%%
\item
For the map \(\nfNmt\) we have to show the other direction
 \[\nfNmt[\Cl(\Cl(\boxTrm[E \cup \SMTeq]) \cup \NOMeq \cup \NMTeq)] \subseteq \Cl(\nfNmt [\boxTrm[E]] \cup \NMTeq)\]
For this it suffices:
\begin{align*}
 &\nfNmt[\Cl(\Cl(\boxTrm[E \cup \SMTeq]) \cup \NOMeq \cup \NMTeq)]\\
 % \subseteq \ &\Cl(\nfNmt[\Cl(\boxTrm[E \cup \SMTeq]) \cup \NOMeq \cup \NMTeq])\\
 % =\ &\Cl(\nfNmt[\Cl(\boxTrm[E \cup \SMTeq])) \cup \nfNmt[\NOMeq \cup \NMTeq])\\
 % \subseteq \ &\Cl(\Cl(\nfNmt[\boxTrm[E \cup \SMTeq]]) \cup \nfNmt[\NOMeq \cup \NMTeq])\\
 % =\ &\Cl(\nfNmt[\boxTrm[E \cup \SMTeq]] \cup \nfNmt[\NOMeq \cup \NMTeq])\\
 \subseteq\ &\Cl(\nfNmt[\boxTrm[E]] \cup \nfNmt[\NOMeq] \cup \nfNmt[\boxTrm[\SMTeq]]  \cup \nfNmt[\NMTeq])\\
 \subseteq \ &\Cl(\nfNmt [\boxTrm[E]] \cup \NMTeq)
 \end{align*}
 % \vspace{-1.6cm}
 % \medskip
 To justify the second inequality, we only need to prove:

 \begin{itemize}
 \item
 \(\nfNmt[\boxTrm[E]] \cup \nfNmt[\NMTeq] \subseteq \Cl(\nfNmt [\boxTrm[E]] \cup \NMTeq)\), which is immediate.
 %%%
 \item
 \(\nfNmt[\boxTrm[\SMTeq]] \subseteq \Cl(\nfNmt [\boxTrm[E]] \cup \NMTeq)\),
which follows since the equations in \(\SMTeq\) get subsumed by \(\NMTeq\) when
 \(\boxTrm\)-ed and normalised via \(\nfNmt\).
 \item
 \(\nfNmt[\NOMeq] \subseteq \Cl(\nfNmt [\boxTrm[E]] \cup \NMTeq)\).
 The only two equations which require any serious verification are~\eqref{NOM-4} and~\eqref{NOM-5}. The proofs of both are essentially
 the same, so we will only consider the first one here:
 \begin{align*}
 \nfNmt([\boldsymbol a \rangle \langle \boldsymbol b |\boldsymbol b'\rangle \hcomp f \langle \boldsymbol c ])
 =\ &\nfNmt([\boldsymbol a \rangle \langle \boldsymbol b |\boldsymbol b'\rangle \langle \boldsymbol x]) \hcomp \nfNmt([\boldsymbol x\rangle f \langle \boldsymbol c ])
 \\
 \eqNMT\ &\nfNmt([\boldsymbol a \rangle \langle \boldsymbol b |\boldsymbol b'\rangle \langle \boldsymbol b']) \hcomp \nfNmt([\boldsymbol b'\rangle f \langle \boldsymbol c ])
 \\
 \eqNMT\ &[\boldsymbol a |\boldsymbol b ] \hcomp [\boldsymbol b |\boldsymbol a ] \hcomp \nfNmt([\boldsymbol a \rangle \langle \boldsymbol b |\boldsymbol b'\rangle \langle \boldsymbol b']) \hcomp \nfNmt([\boldsymbol b'\rangle f \langle \boldsymbol c ])
 \\
 \eqNMT\ &[\boldsymbol a |\boldsymbol b ] \hcomp \nfNmt([\boldsymbol b \rangle \langle \boldsymbol b |\boldsymbol b'\rangle \langle \boldsymbol b']) \hcomp \nfNmt([\boldsymbol b'\rangle f \langle \boldsymbol c ])
 \\
 \eqNMT\ &[\boldsymbol a |\boldsymbol b ] \hcomp \nfNmt([\boldsymbol b'\rangle f \langle \boldsymbol c ])
 \end{align*}

 % For the justification of \(\nfNmt([\boldsymbol b \rangle \langle \boldsymbol b |\boldsymbol b'\rangle \langle \boldsymbol b']) \eqNMT id\) see the remark below.
 \end{itemize}
\end{itemize}

\noindent
Finally, we show that the maps \(\nfNmt\) and \(\injNmt\) are inverses of
each other.
We have \(\nfNmt \circ \injNmt\,(f) = f\) for any \(f \in \nTrm(\Sigma )\), by induction on \(f\).
We have {\normalfont \(\injNmt \circ \nfNmt\,(f) \eqNOM f\)} for any
\(f \in \nTrm(\Trm(\Sigma ))\), where
{\normalfont \[\label{eqNOM-def}\,\eqNOM\,\]}is
equality up to the equations \(\NOMeq \cup \NMTeq \cup \boxTrm[\SMTeq]\), also by induction on $f$.
The interesting case is \(f = [ab\rangle \sigma \langle cd]\) for the twist $\sigma$:\begin{align*}
\injNmt \circ \nfNmt\,([ab\rangle \Sigma \langle cd])\
&\ =\ \, \injNmt \,([ab|dc])\\
&\ =\ \ [ab|dc]\ \eqNOM\ [ba|cd]\ \eqNOM\ [ba\rangle id \langle cd] \\
&\eqNOM\ [ab|ab] \hcomp [ba\rangle id \langle cd] \\
&\eqNOM\ [ab\rangle \langle ab|ba\rangle \hcomp id \langle cd] \\
&\eqNOM\ [ab\rangle \langle ab|ba\rangle \langle cd] \\
&{\ \  =\ \ \hspace{.25em}}[ab\rangle \Sigma \langle cd]
\end{align*}
% For full details we refer to~\cite{balco:phd}.

This finishes the proof.
\end{proof}
\end{lem}

In Section~\ref{sec:examples}, we introduced the nominal monoidal theories for the categories
of bijections, injections, surjections and functions (amongst
others), see \autoref{fig:nmt-theories}. We show now how to prove the completeness of these theories  from the completeness of the corresponding $\SMT$s of \autoref{fig:smt-theories}.

Recall that completeness of an \(\SMT\) \(\langle \Sigma ,E\rangle \) with regards to some
category \(\mathbb C\) means that the \(\PROP\) presented by
\(\langle \Sigma ,E\rangle \) is isomorphic to \(\mathbb C\) \[\Prop\,\langle \Sigma ,E\rangle \cong \mathbb C.\]
Likewise, the completeness of an \(\NMT\) with regards to
some category \(\mathsf n\mathbb C\) is the existence of an isomorphism
\[\nProp\,\langle \Sigma ,E\rangle \cong \mathsf n\mathbb C\]
\begin{thm}[Completeness of \(\NMT\)s]\label{thm:NMTcompleteness}
The calculi of \autoref{fig:nmt-theories} are sound and complete, that is, the
categories presented by these calculi are isomorphic to the categories
of finite sets of names with the respective maps.

\begin{proof}
We show the result for the category of finite functions \(\nF\). Similar
arguments apply to the other theories presented in \autoref{fig:nmt-theories}.
 In order to show completeness of
the nominal theory of functions w.r.t. \(\nF\), we start with the
\(\SMT\) \(\langle \Sigma_{\FunSkel} ,E_{\FunSkel} \rangle \) of \autoref{fig:smt-theories}.
We know
\[\NOM(\Prop\,\langle \Sigma_{\FunSkel} ,E_{\FunSkel} \rangle ) \cong \nProp(\Nmt\,\langle \Sigma_{\FunSkel} ,E_{\FunSkel} \rangle )\]
from Lemma~\ref{lem:completeSquareNMT},
\[\NOM(\FunSkel)\cong \nF\]
from Example~\ref{ex:nFNOMiso} and
\[\Prop\,\langle \Sigma_{\FunSkel} ,E_{\FunSkel} \rangle \cong \FunSkel\]
from completeness of \(\langle \Sigma_{\FunSkel} ,E_{\FunSkel} \rangle \)
for \(\FunSkel \).
Putting these together, we
obtain \[\nProp(\Nmt\,\langle \Sigma_{\FunSkel} ,E_{\FunSkel} \rangle )\cong \nF\] that is, \(\Nmt\,\langle \Sigma_{\FunSkel} ,E_{\FunSkel} \rangle \) is
complete for \(\nF\).
\end{proof}
\end{thm}

\hypertarget{sec:nfSmt}{%
\subsection{Translating NMTs into SMTs}\label{sec:nfSmt}}

This section follows the same lines as Section~\ref{sec:nfNmt}, but now translating nominal monoidal theories into symmetric monoidal theories. Indeed, we can convert an $\NMT$ into an $\SMT$ by first embedding nominal equations into ordinary string diagrams and then normalising the diagrams via a function $\nfSmt$, which we are going to define now.

Compared to what we have seen when we defined $\nfNmt$, normalising embedded nominal string diagrams into ordinary string diagrams is slightly more tricky. This is due to the fact that in nominal sequential composition, we are allowed to compose two diagrams which share the same set of output and input labels, disregarding the order of the named ports.

For example, in the picture below, we see a wire crossing inside the purple box, introduced by the fact that the ports of the box interface and the ports of the generator inside the box have to be lined up.

\begin{center}
\includegraphics[page=69, width=7cm]{twists_new_thesis}
\end{center}

However, no such crossing is (directly) visible in the one-dimensional syntax ${\color{magenta}\langle a,b]}{\color{cyan}[b,a\rangle \mu \langle c]}{\color{magenta}[c \rangle}$. Thus, when translating such a diagram back into ordinary string diagrams, we might need to insert some symmetries, i.e.\  the diagram
${\color{magenta}\langle a,b]}{\color{cyan}[b,a\rangle \mu \langle c]}{\color{magenta}[c \rangle}$ should normalise to $\sigma \hcomp \mu$:

\begin{center}
\includegraphics[page=70, width=7cm]{twists_new_thesis}
\end{center}

The purple box in the pictures above orders sets of labelled wires. We write it as $\diaTrm$ in one-dimensional notation and define it as follows. Recall that given a set $A$ of names, we write $\boldsymbol a$ for a list that fixes an enumeration of the elements of $A$. The notation  $\langle \boldsymbol a]\,t\,[\boldsymbol b\rangle$ is the one from Proposition~\ref{prop:ORD}.

\begin{itemize}
\item
 \(\label{dia-def}\diaTrm(t : A \to B) = \langle \boldsymbol a]\,t\,[\boldsymbol b\rangle \)
 with \(set(\boldsymbol a) = A\) and \(set(\boldsymbol b) = B\) and
\item
 \(\diaTrm(E) = \{ \langle \boldsymbol a]\,s\,[\boldsymbol b\rangle = \langle \boldsymbol a]\,t\,[\boldsymbol b\rangle \mid s=t \in E\}\)
\end{itemize}

\noindent
Now, we can define
\[\label{nfSmt-def}\nfSmt : \Trm(\diaTrm(\nTrm(\Sigma ))) \to \Trm(\Sigma )\]
as
\begin{align*}& \nfSmt\, (\langle \boldsymbol a ][\boldsymbol a'\rangle \gamma \langle \boldsymbol b'][\boldsymbol b \rangle ) = \langle \boldsymbol a |\boldsymbol a'\rangle \hcomp \underline{\gamma } \hcomp \langle \boldsymbol b'|\boldsymbol b \rangle \text{ where } \gamma \in \Sigma \\
 & \nfSmt\, (\langle a] id_a [a\rangle ) = id\\
 & \nfSmt\, (\langle a] \delta_{ab} [b\rangle ) = id\\
 & \nfSmt\, (\langle \boldsymbol a ] f \hcomp g [\boldsymbol c \rangle ) = \nfSmt\, (\langle \boldsymbol a ] f [\boldsymbol b \rangle ) \hcomp \nfSmt\, (\langle \boldsymbol b ] g [\boldsymbol c \rangle )\\
 & \nfSmt\, (\langle \boldsymbol a ] f \uplus g [\boldsymbol b \rangle ) = \langle \boldsymbol a |\boldsymbol a_1\concat \boldsymbol a_2\rangle \hcomp (\nfSmt\, (\langle \boldsymbol a_1] f [\boldsymbol b_1\rangle ) \oplus \nfSmt\, (\langle \boldsymbol a_2] g[\boldsymbol b_2\rangle )) \hcomp \langle \boldsymbol b_1\concat \boldsymbol b_2|\boldsymbol b \rangle \\
 & \nfSmt\, (\langle \boldsymbol a ] \pi \cdot f [\boldsymbol b \rangle ) = \nfSmt\, (\langle \pi^{-1} \cdot \boldsymbol a ]\, f \,[\pi^{-1}\cdot \boldsymbol b \rangle ) \\
 & \nfSmt\, (\gamma ) = \gamma \text{ where } \gamma \in \Sigma \\
 & \nfSmt\, (id) = id\\
 & \nfSmt\, (\sigma ) = \sigma \\
 & \nfSmt\, (f \hcomp g) = \nfSmt\, (f) \hcomp \nfSmt\, (g)\\
 & \nfSmt\, (f \oplus g) = \nfSmt\, (f) \oplus \nfSmt\, (g)\\
\end{align*}

\begin{defi}\label{Smt-def}
We define \(\Smt : \NMT \to \SMT\) as
\[\Smt\,\langle \Sigma ,E\rangle = \langle \Sigma ,\nfSmt(\diaTrm(E))\rangle .\]
\end{defi}

In terms of Figure~\ref{fig:smt-compl-sq}, the definition above corresponds to going around the square down-right. To prepare us for the next subsection, we also recall how to go around right-down. This is the converse to the construction at the start of Section~\ref{sec:nom-translation}, now taking an \(\NMT\) \(\langle \Sigma , E\rangle \) to
\(\langle \Trm(\diaTrm(\nTrm(\Sigma ))), \diaTrm(E) \cup \ORDeq\rangle \), where  \(\label{dia-def}\diaTrm\) is as above and \(\ORDeq\) are the equations from Proposition~\ref{prop:ORD}. In the next section, we show that this construction and the one in Definition~\ref{Smt-def} give rise to the same $\PROP$s.

\hypertarget{completeness-of-mathsfsmts}{%
\subsection{Completeness of SMTs}\label{completeness-of-mathsfsmts}\label{sec:SMTcompleteness}}

While Section~\ref{sec:NMTcompleteness} showed how to transfer completeness of $\SMT$s to completeness of $\NMT$s, we now go into the opposite direction. Indeed, we show how to prove the completeness of the $\Smt(\Sigma,E)$ from the completeness of a nominal monoidal theory $\langle \Sigma,E\rangle$.
The central observation is that the diagram in \autoref{fig:smt-compl-sq} commutes up to isomorphism.

\begin{figure}
\[\xymatrix{
% Level 1
\NMT
\ar[0,4]|-{\ \nProp\ }
\ar[5,0]_{\SmtSubscript}
& & & &
\nPROP
\ar[5,0]^{\ORDSubscript}
\\
% Level 2
& \langle \Sigma , E\rangle
\ar@{ |->}[0,2]
\ar@{ /->}`d[dddr][dddrr]
& & \nProp\, \langle \Sigma , E\rangle
\ar@{ /->}[dd]
 &
\\
% Level 3
 & & & &
\\
% Level 4
 & & &
\ORD(\nProp\, \langle \Sigma , E\rangle )
\ar@/^1.2pc/[d]^{\nfSmtSubscript}
\\
% Level 5
 & & &
\Prop(\Smt\, \langle \Sigma , E\rangle )
\ar@/^1.2pc/[u]^{\injSmtSubscript}
\\
% Level 6
\SMT
\ar[0,4]|-{\ \Prop\ }
& & & &
\PROP
}\]
\caption{Completing the square}\label{fig:smt-compl-sq}
\end{figure}

First, we define the homomorphism of term-algebras
\(\label{injSmt-def} \injSmt : \Trm(\Sigma ) \to \Trm(\diaTrm(\nTrm(\Sigma )))\):
\begin{align*}& \injSmt(\underline{\gamma }) = \langle \boldsymbol a ][\boldsymbol a \rangle \gamma \langle \boldsymbol b ][\boldsymbol b \rangle \text{ where } \gamma \in \Sigma \\
 & \injSmt(id) = id\\
 & \injSmt(\Sigma ) = \Sigma \\
 & \injSmt(f \hcomp g) = \injSmt(f) \hcomp \injSmt(g)\\
 & \injSmt(f \oplus g) = \injSmt(f) \oplus \injSmt(g)\\
\end{align*}

Next, we show that \(\nfSmt\) and \(\injSmt\) are inverses of
each other.

\begin{lem}\label{lem:completeSquareSMT}
The diagram in \autoref{fig:smt-compl-sq} commutes up to isomorphism.
\begin{proof}
We want to show that the two maps \(\nfSmt\) and \(\injSmt\) are
isomorphisms. By definition, both \(\nfSmt\) and \(\injSmt\) are
homomorphisms between the term algebras.
 % and we have shown in
% Lemma~\ref{lem:nfSmtInjEq} that \(\nfSmt \circ \injSmt\,(f) \eqSMT f\) and
% \(\injSmt \circ \nfSmt\,(f) \eqORD f\) follows from Lemma~\ref{lem:injSmtNfEq}.
To verify that these maps are well-defined, that is, that they respect
equivalence of \(\Trm\)s, we need to check that they preserve
the following equations:

\begin{itemize}
\item
For the map \(\injSmt\), we have to show
 \[\injSmt[\Cl(\nfSmt [\diaTrm[E]] \cup \SMTeq)] \subseteq \Cl(\Cl(\diaTrm[E \cup \NMTeq]) \cup \ORDeq \cup \SMTeq)\]
 In fact, by Lemma~\ref{lem:nfNmtCl}, it suffices to check that
 \[\injSmt[\nfSmt [\diaTrm[E]]] \subseteq \Cl(\diaTrm[E \cup \diaTrm[\NMTeq] \cup \ORDeq \cup \SMTeq)\]
 and \(\injSmt[\SMTeq] \subseteq \Cl(\SMTeq)\). The first inequality follows
 from
 \(\injSmt[\nfSmt [\diaTrm[E]] \eqORD \diaTrm[E]\), where $\label{eqORD-def}\,\eqORD\,$ is equality up to the equations $\ORDeq \cup \SMTeq \cup \diaTrm[\NMTeq]$. The second
 is immediate.
\item
For the map \(\nfSmt\), we have to show the other direction
 \[\nfSmt[\Cl(\Cl(\diaTrm[E \cup \NMTeq]) \cup \ORDeq \cup \SMTeq)] \subseteq \Cl(\nfSmt [\diaTrm[E]] \cup \SMTeq)\]
for which it suffices:
\begin{align*}
&\nfSmt[\Cl(\Cl(\diaTrm[E \cup \NMTeq]) \cup \ORDeq \cup \SMTeq)]\\
% \subseteq \ &\Cl(\nfNmt[\Cl(\diaTrm[E \cup \NMTeq]) \cup \ORDeq \cup \SMTeq])\\
% =\ &\Cl(\nfNmt[\Cl(\diaTrm[E \cup \NMTeq])) \cup \nfNmt[\ORDeq \cup \SMTeq])\\
% \subseteq \ &\Cl(\Cl(\nfSmt[\diaTrm[E \cup \NMTeq]]) \cup \nfSmt[\ORDeq \cup \SMTeq])\\
% =\ &\Cl(\nfSmt[\diaTrm[E \cup \NMTeq]] \cup \nfSmt[\ORDeq \cup \SMTeq])\\
\subseteq \ &\Cl(\nfNmt[\diaTrm[E]]  \cup \nfSmt[\SMTeq] \cup \nfSmt[\diaTrm[\NMTeq]] \cup \nfSmt[\ORDeq])\\
\subseteq \ &\Cl(\nfSmt [\diaTrm[E]] \cup \SMTeq)
\end{align*}
To justify the last inequality, we show:
 \begin{itemize}
 \item
 \(\nfSmt[\diaTrm[E]] \cup \nfSmt[\SMTeq] \subseteq \Cl(\nfSmt [\diaTrm[E]] \cup \SMTeq)\), which is immediate.
 %%%
 \item
 \(\nfSmt[\diaTrm[\NMTeq]] \subseteq \Cl(\nfSmt [\diaTrm[E]] \cup \SMTeq)\)
 It is easy enough to see that most equations of
 \(\nfSmt[\diaTrm[\NMTeq]]\) are in \(\Cl(\SMTeq)\). For the
 interesting case of~\eqref{NMT-comm} being preserved by
 \(\nfSmt \circ \diaTrm\), see the proof of~\cite[Proposition 7.25]{balco:phd}.
 \item
 \(\nfSmt[\ORDeq] \subseteq \Cl(\nfSmt [\diaTrm[E]] \cup \SMTeq)\).
 The only two equations which require any serious verification are~\eqref{ORD-4} and~\eqref{ORD-5}. The proofs of both are essentially
 the same, so we will only consider the first one here:
\begin{align*}
 \nfSmt(\langle \boldsymbol a ] [\boldsymbol a'|\boldsymbol b ] ; f [\boldsymbol c \rangle )
 =\ &\nfSmt(\langle \boldsymbol a ][\boldsymbol a'|\boldsymbol b ][\boldsymbol b'\rangle ) \hcomp \nfSmt(\langle \boldsymbol b'] f [\boldsymbol c \rangle )
 \\
 \eqSMT\ &\nfSmt(\langle \boldsymbol a ][\boldsymbol a'|\boldsymbol b ][\boldsymbol b \rangle ) \hcomp \nfSmt(\langle \boldsymbol b ] f [\boldsymbol c \rangle )
 \\
 \eqSMT\ &\langle \boldsymbol a |\boldsymbol a'\rangle \hcomp \nfSmt(\langle \boldsymbol b ] f [\boldsymbol c \rangle )
 \\
 =\ &\nfSmt(\langle \boldsymbol a |\boldsymbol a'\rangle ) \hcomp \nfSmt(\langle \boldsymbol b ] f [\boldsymbol c \rangle )
 \\
 =\ &\nfSmt(\langle \boldsymbol a |\boldsymbol a'\rangle \hcomp \langle \boldsymbol b ] f [\boldsymbol c \rangle )
 \end{align*}

 For these equalities to hold, we need to show
 \[\nfSmt(\langle \boldsymbol a ]t[\boldsymbol b'\rangle ) \hcomp \langle \boldsymbol b'|\boldsymbol b \rangle \eqSMT \nfSmt(\langle \boldsymbol a ]t[\boldsymbol b \rangle )\] which
 follows by induction on \(t\).
 \end{itemize}
\end{itemize}

\noindent
To show that $\nfSmt$ and $\injSmt$ are inverses, we have \(\nfSmt \circ \injSmt\,(f) \eqSMT f\) for any \(f \in \Trm(\Sigma )\), by induction on \(f\). The only case of interest is
\(f = \underline{\gamma }\) where \(\gamma \in \Sigma \):
\[\nfSmt \circ \injSmt\,(\underline{\gamma }) = \nfSmt\,(\langle \boldsymbol a ][\boldsymbol a \rangle \gamma \langle \boldsymbol b ][\boldsymbol b \rangle ) = \langle \boldsymbol a |\boldsymbol a \rangle \hcomp \underline{\gamma } \hcomp \langle \boldsymbol b |\boldsymbol b \rangle \eqSMT \underline{\gamma }\]
Finally, we have \(\injSmt \circ \nfSmt\,(f) \eqORD f\) for any
\(f \in \Trm(\diaTrm(\nTrm(\Sigma )))\).
\end{proof}
\end{lem}
To conclude this section, we give a result analogous to
Theorem~\ref{thm:NMTcompleteness}.

\begin{thm}[Completeness of \(\SMT\)s]\label{thm:SMTcompleteness}
If the \(\NMT\) \(\langle \Sigma ,E\rangle \)  is complete for some
\(\mathsf n\mathbb C\) and \(\ORD(\mathsf n\mathbb C) \cong \mathbb C\),
then \(\Smt\,\langle \Sigma_{\FunSkel} ,E_{\FunSkel} \rangle \) is complete for \(\mathbb C\).

\begin{proof}
From Lemma~\ref{lem:completeSquareSMT} we know that
\[\ORD(\nProp\,\langle \Sigma ,E\rangle ) \cong \Prop(\Smt\,\langle \Sigma ,E\rangle )\] From completeness of
\(\langle \Sigma ,E\rangle \) for \(\mathsf n\mathbb C\) we know
\[\nProp\,\langle \Sigma ,E\rangle \cong \mathsf n\mathbb C\] Putting these together, we obtain
\[\Prop(\Smt\,\langle \Sigma ,E\rangle )\cong \mathbb C \qedhere\]
\end{proof}
\end{thm}

The theorem above can be useful when trying to prove soundness and
completeness of an ordinary \(\SMT\), where the presented \(\NMT\) is
easier to prove sound and complete, like in the case of bijections described in Section~\ref{sec:examples}:

\begin{exa}
The completeness proof for ordinary string diagrams given in~\cite{lafont} is already quite complex in the case of bijections because it requires a complete calculus for permutations. On the other hand, the corresponding nominal theory of bijections has a trivial completeness proof because all nominal string diagrams in the theory of bijections normalise to a disjoint union of renamings. We can now apply the theorem above to obtain an easy completeness proof for the corresponding symmetric monoidal theory of bijections.
\end{exa}

% !TEX root =  calco2019-lmcs.tex

\section{Conclusion}

The equivalence of nominal and ordinary PROPs (Theorem~\ref{thm:equivalence}), as well as the equivalence of nominal and symmetric monoidal theories (Theorems~\ref{thm:SMTcompleteness} and~\ref{thm:NMTcompleteness})  has a satisfactory graphical interpretation. Indeed, comparing \autoref{fig:smt-theories} and \autoref{fig:nmt-theories}, truncated and shown side by side below, we see that both share, modulo different labellings of wires mediated by the functors $\ORD$ and $\NOM$, the same core of generators and equations. The difference lies only in the equations expressing, on the one hand, that $\oplus$ has natural symmetries and, on the other hand, that generators are a nominal set and that $\uplus$ is commutative. In fact, this can be taken as a justification of the importance of the naturality of symmetries, which, informally speaking, compensates for the irrelevant detail induced by ordering names.

\begin{center}
\includegraphics[page=66, width=\linewidth]{twists_new_thesis}
\end{center}

There are several directions for future research. First, the notion of an internal monoidal category has been developed because it is easier to prove the basic results in general rather than only in the special case of nominal sets. Nevertheless, it would be interesting to explore whether there are more interesting instances of internal monoidal categories in other semi-cartesian closed categories.

Second, internal monoidal categories are a principled way to build monoidal categories with a partial tensor. By working internally in the category of nominal sets with the separated product we can capture in a natural way constraints such as the tensor $f\oplus g$ for two partial maps $f,g:\names\to V$ being defined only if the domains of $f$ and $g$ are disjoint. This reminds us of the work initiated by O'Hearn and Pym on categorical and algebraic models for separation logic and other resource logics, see e.g.~\cite{ohearn-pym,galmiche-etal,struth}. It seems promising to investigate how to build categorical models for resource logics based on internal monoidal theories. In one direction, one could extend the work of Curien and Mimram~\cite{curien-mimram} to partial monoidal categories.

Third, there has been substantial progress in exploiting Lack's work on composing PROPs \cite{lack} in order to develop novel string diagrammatic calculi for a wide range of applications, see e.g.~\cite{rewriting-modulo,signal-flow-1}. It will be interesting to explore how much of this technology can be transferred from PROPs to nominal PROPs. % chktex 2

Fourth, various applications of nominal string diagrams could be of interest. The original motivation for our work was to obtain a convenient calculus for simultaneous substitutions that can be integrated with multi-type display calculi~\cite{Multitype} and, in particular, with the multi-type display calculus for first-order logic of Tzimoulis~\cite{tzimoulis}. Another direction for applications comes from the work of Ghica and Lopez~\cite{ghica-lopez} on a nominal syntax for string diagrams. In particular, it would be of interest to add various binding operations to nominal PROPs.

More generally, we expect nominal PROPs to play a role as an intermediate level of abstraction for the implementation of programming languages that have a denotational semantics as string diagrams. As case study, it would be interesting to look at the Bayesian Networks of~\cite{jacobs-zanasi,jkz}. % which provide a syntax for networks that has a formal semantics as string diagrams.

%%%%%%%%%%%%
\bibliography{calco2019}
\bibliographystyle{alphaurl}

% !TEX root =  calco2019-lmcs.tex

\appendix
\section{A review of internal category theory}%
\label{app:internalCats}

We review the notation that we use in Section~\ref{sec:internal-monoidal}. We consulted Borceux, Handbook of Categorical Algebra, Volume 1, Chapter 8 and the \href{https://ncatlab.org/nlab/show/internal+category}{nLab}, adapting the notation to our needs.

\begin{defi}[internal category]%
\label{def:internal-cat}
In a category with finite limits an \emph{internal category}  is a diagram
\begin{equation}\label{eq:def-internal-category}
\vcenter{
\xymatrix@C=12ex{
A_3 \
\ar@<3ex>[0,1]|-{\ \textit{right}\ }
\ar@<1ex>[0,1]|-{\ \textit{compr}\ }
\ar@<-1ex>[0,1]|-{\ \textit{compl}\ }
\ar@<-3ex>[0,1]|-{\ \textit{left}\ }
&
{\ \ A_2 \ \ }
\ar@<2ex>[0,1]|-{\pi_2}
\ar[0,1]|-{\ \comp\ }
\ar@<-2ex>[0,1]|-{\pi_1}
&
{\ \ A_1 \ \ } %\ar@(ur,ul)_{s}
%\ar@<2ex>[0,1]^-{d^1_1}
\ar@<2ex>[0,1]|-{\ \dom\ }
%\ar@<-2ex>[0,1]_-{d^1_0}
\ar@<-2ex>[0,1]|-{\ \cod\ }
&
\ A_0
\ar[0,-1]|-{\ i \ }
}
}
\end{equation}
where
\begin{enumerate}
\item the diagram \  \
$\vcenter{
\xymatrix{
A_2
\ar[r]^{\pi_2}
\ar[d]_{\pi_1}
&
A_1\ar[d]^{\dom} \\
A_1\ar[r]^{\cod}
&
A_0
}
}$
\ \ is a pullback,
\item $\dom\circ \comp=\dom\circ \pi_1$ \ \ and \ $\cod\circ
  \comp=\cod\circ \pi_2$,
\item $\dom\circ i  = \id_{A_0} = \cod\circ i$,
\item $\comp\circ \langle i\circ\dom,\id_{A_1} \rangle = \id_{A_1} = \comp\circ \langle\id_{A_1},i\circ\cod \rangle$
\item $\comp\,\circ\,\compl= \comp\,\circ\,\compr$
\end{enumerate}
where
\begin{itemize}
\item $\langle i\circ\dom,\id_{A_1}\rangle :A_1\to A_2$ and $\langle \id_{A_1},i\circ\cod\rangle :A_1\to A_2$ are the arrows into the pullback $A_2$  pairing $i\circ\dom,\id_{A_1}:A_1\to A_1$ and $\id_{A_1},i\circ\cod:A_1\to A_1$, respectively;
\item the ``triple of arrows''-object $A_3$ is the pullback
\[\vcenter{
\xymatrix{
A_3
\ar[r]^{\textit{right}}
\ar[d]_{\textit{left}}
&
A_2\ar[d]^{\pi_1} \\
A_2\ar[r]^{\pi_2}
&
A_1
}
}\]
where, intuitively, $\textit{left}$ ``projects out the left two arrows'' and $\textit{right}$ ``projects out the right two arrows'';
\item $\compl$ is the arrow composing the ``left two arrows''
\[\vcenter{
\xymatrix@C=10ex{
A_3
\ar@/^/[rrd]^{\ \ \pi_2\,\circ\, \textit{right}}
\ar@{..>}[dr]|-{\compl}
\ar@/_/[rdd]_{\comp\,\circ\, \textit{left}}
&&\\
&
A_2
\ar[r]^{\pi_2}
\ar[d]_{\pi_1}
&
A_1\ar[d]^{\dom}
\\
&
A_1\ar[r]^{\cod}
&
A_0
}
}\]
\item $\compr$ is the arrow composing the ``right two arrows''
\[\vcenter{
\xymatrix@C=10ex{
A_3
\ar@/^/[rrd]^{\ \ \comp\,\circ\, \textit{right}}
\ar@{..>}[dr]|-{\compr}
\ar@/_/[rdd]_{\pi_1\,\circ\, \textit{left}}
&&\\
&
A_2
\ar[r]^{\pi_2}
\ar[d]_{\pi_1}
&
A_1\ar[d]^{\dom}
\\
&
A_1\ar[r]^{\cod}
&
A_0
}
}\]
\end{itemize}
\end{defi}

\begin{rem}
1.~and 2.~define $A_2$ as the `object of composable pairs of arrows' while 3.~and 4.~express that the `object  of arrows' $A_1$ has identities and 5.~formalises associativity of composition. Since $A_2$ and $A_3$ are pullbacks, the structure is defined completely by $(A_0,A_1, \dom, \cod, i, \comp)$ only, but including $A_3$ as well as $\compr,\compl,\textit{right},\textit{left}, \pi_2,\pi_1$ helps writing out the equations.
\end{rem}

\begin{defi}\label{def:internal-functor}
A morphism $f:A\to B$ between internal categories, an \emph{internal functor}, is a pair $(f_0,f_1)$ of arrows such that the six squares (one for each of $\pi_2$, $\comp$, $\pi_1$, $\dom$, $\cod$, $i$)
\begin{equation}\label{eq:def:internal-functor}
\vcenter{
\xymatrix@C=12ex{
A_2 \
\ar@<2ex>[0,1]|-{\pi_2}
\ar[0,1]|-{\ \comp\ }
\ar@<-2ex>[0,1]|-{\pi_1}
\ar[dd]_{f_2}
&
{\ \ A_1 \ \ } %\ar@(ur,ul)_{s}
\ar@<2ex>[0,1]|-{\ \dom\ }
\ar@<-2ex>[0,1]|-{\ \cod\ }
\ar[dd]_{f_1}
&
\ A_0
\ar[0,-1]|-{\ i\ }
\ar[dd]_{f_0}
\\
&&
\\
B_2 \
%\ar@<2ex>[0,1]^-{d^2_2}
%\ar[0,1]|-{d^2_1}
\ar@<2ex>[0,1]|-{\pi_2}
\ar[0,1]|-{\ \comp\ }
\ar@<-2ex>[0,1]|-{\pi_1}
%\ar@<-2ex>[0,1]_-{d^2_0}
&
{\ \ B_1 \ \ } %\ar@(ur,ul)_{s}
\ar@<2ex>[0,1]|-{\ \dom\ }
\ar@<-2ex>[0,1]|-{\ \cod\ }
&
\ B_0
\ar[0,-1]|-{\ i\ }
}
}
\end{equation}
commute.
\end{defi}

\begin{rem}
\begin{itemize}
\item
Because $B_2$ is a pullback $f_2$ is uniquely determined by $f_1$.  In more detail, if $\Gamma\to B_2$ is any arrow then, because $B_2$ is a pullback, it can be written as a pair
\begin{equation}\label{eq:pairing}
\langle l,r\rangle:\Gamma\to B_2
\end{equation}
of arrows $l,r:\Gamma\to B_1$ and $f_2$ is determined by $f_1$ via
\begin{equation}\label{equ:f2}
f_2\circ\langle l,r\rangle=\langle f_1\circ l,f_1\circ r\rangle
\end{equation}
%%%
\item Even if $f_2$ is not needed as part of the structure in the above definition, including $f_2$ makes it easier to state that $f_1$ preserves composition.
%%%
\item Similarly, $B_3$ is a pullback, and there is a unique arrow $f_3$ such that $(f_0,f_1,f_2,f_3)$ together make further 4 squares commute, one for each of $\textit{right}, \compr, \compl,\textit{left}$, see~\eqref{eq:def-internal-category}. We may include $f_3$ in the structure whenever convenient.
\end{itemize}
\end{rem}

\begin{defi}
A natural transformation $\alpha:f\to g$ between internal functors $f,g:A\to B$, an \emph{internal natural transformation},   is an arrow $\alpha:A_0\to B_1$ such that, recalling~\eqref{eq:pairing},
\[
\dom\circ\alpha = f_0
\quad\quad
\cod\circ\alpha = g_0
\quad\quad
\comp\circ\langle f_1,\alpha\circ\cod\rangle=\comp\circ\langle\alpha\circ\dom,g_1\rangle
\]
\end{defi}

\begin{rem}
Internal categories with functors and natural transformations form a 2-category. We denote by $\Cat(\mathcal V)$ the category or 2-category of categories internal in $\mathcal V$. The forgetful functor $\Cat(\mathcal V)\to\mathcal V$ mapping an internal category $A$ to its object of objects $A_0$ has both left and right adjoints and, therefore, preserves limits and colimits. Moreover, a limit of internal categories is computed component-wise as $(\lim D)_j=\lim (D_j)$ for $j=0,1,2$.
\end{rem}

\begin{rem}
A monoidal category can be thought of both as a monoid in the category of categories and as an internal category in the category of monoids. To understand this in more detail, note that both definitions give rise to the diagram
\[
\xymatrix@C=20ex{
A_2\times A_2 \
\ar[0,1]|-{\ \comp\times \comp\ }
\ar[dd]_{m_2}
&
{\ \ A_1\times A_1 \ \ }
\ar@<1ex>[0,1]^-{\dom\times \dom}
\ar@<-1ex>[0,1]_-{\cod\times \cod}
\ar[dd]_{m_1}
&
\ A_0\times A_0
\ar[dd]_{m_0}
\\
&&
\\
A_2 \
\ar[0,1]|-{\ \comp\ }
&
{\ \ A_1 \ \ }
\ar@<1ex>[0,1]^-{\dom}
\ar@<-1ex>[0,1]_-{\cod}
&
\ A_0
}
\]
where
\begin{itemize}
\item in the case of a monoid $A$ in the category of internal categories, $m=(m_0,m_1,m_2)$ is an internal functor $A\times A \to A$ and, using that products of internal categories are computed component-wise, we have $\comp\circ m_2=m_1\circ (\comp\times \comp)$, which gives us the interchange law \[(f\then\, g)\cdot(f'\then\, g')= (f\cdot f')\,\then\,(g\cdot g')\]
by using~\eqref{equ:f2} with $m$ for $f$ and  writing $\then$ for $\comp$ and  $\cdot$ for $m_1$;
\item in the case of a category internal in monoids we have monoids $A_0,A_1,A_2$ and monoid homomorphisms $i,\dom,\cod,\comp$ which, if spelled out, leads to the same commuting diagrams as the previous item.
\end{itemize}
\end{rem}

\begin{rem}
In Section~\ref{sec:internal-monoidal} we answer the question of how to internalise the cartesian product $A\times A$ in a monoidal category $(\mathcal V,I,\otimes)$. In other words, where above we use $A_0\times A_0$ we now want $A_0\otimes A_0$. This rules out the second item above, leaving only the definition of a monoidal category as a monoid in the category of internal categories.
\end{rem}

\end{document}